**Ultra-Sharp Nanowire Arrays Natively Permeate, Record, and Stimulate Intracellular Activity in Neuronal and Cardiac Networks**

*Ren Liu†, Jihwan Lee†, Youngbin Tchoe†, Deborah Pre, Andrew M. Bourhis, Agnieszka D'Antonio-Chronowska, Gaelle Robin, Sang Heon Lee, Yun Goo Ro, Ritwik Vatsyayan, Karen J. Tonsfeldt, Lorraine A. Hossain, M. Lisa Phipps, Jinkyoung Yoo, John Nogan, Jennifer S. Martinez, Kelly A. Frazer, Anne G. Bang, Shadi A. Dayeh**

† These authors contributed equally.

Dr. R. Liu, J. Lee, Dr. Y. Tchoe, A. M. Bourhis, Dr. S. H. Lee, Dr. Y. G. Ro, R. Vatsyayan, Dr. K. J. Tonsfeldt, Dr. L. A. Hossain, Prof. S. A. Dayeh
Integrated Electronics and Biointerfaces Laboratory
Department of Electrical and Computer Engineering
University of California San Diego
9500 Gilman Drive, La Jolla, CA 92093, USA
E-mail: sdayeh@eng.ucsd.edu

Dr. D. Pre, Dr. G. Robin, Prof. A. G. Bang
Conrad Prebys Center for Chemical Genomics
Sanford Burnham Prebys Medical Discovery Institute
10901 North Torrey Pines Road, La Jolla, CA 92037, USA

Dr. A. D'Antonio-Chronowska, Prof. K. A. Frazer
Department of Pediatrics
University of California San Diego,
9500 Gilman Drive, La Jolla, CA 92093, USA

Dr. K. J. Tonsfeldt
Center for Reproductive Science and Medicine
Department of Obstetrics, Gynecology, and Reproductive Sciences
University of California San Diego
9500 Gilman Drive, La Jolla, CA 92093, USA

Dr. M. L. Phipps, Dr. J. Yoo
Center for Integrated Nanotechnologies
Los Alamos National Laboratory
Los Alamos, NM 87545, USA

Dr. J. Nogan
Center for Integrated Nanotechnologies
Sandia National Laboratories
Albuquerque, NM 87185, USA

Prof. J. S. Martinez

Center for Materials Interfaces in Research and Applications and Department of Applied Physics and Materials Science
Northern Arizona University
624 S. Knoles Dr. Flagstaff, AZ 86011

Intracellular access with high spatiotemporal resolution can enhance our understanding of how neurons or cardiomyocytes regulate and orchestrate network activity, and how this activity can be affected with pharmacology or other interventional modalities. Nanoscale devices often employ electroporation to transiently permeate the cell membrane and record intracellular potentials, which tend to decrease rapidly to extracellular potential amplitudes with time. Here, we report innovative scalable, vertical, ultra-sharp nanowire arrays that are individually addressable to enable long-term, native recordings of intracellular potentials. We report large action potential amplitudes that are indicative of intracellular access from 3D tissue-like networks of neurons and cardiomyocytes across recording days and that do not decrease to extracellular amplitudes for the duration of the recording of several minutes. Our findings are validated with cross-sectional microscopy, pharmacology, and electrical interventions. Our experiments and simulations demonstrate that individual electrical addressability of nanowires is necessary for high-fidelity intracellular electrophysiological recordings. This study advances our understanding of and control over high-quality multi-channel intracellular recordings, and paves the way toward predictive, high-throughput, and low-cost electrophysiological drug screening platforms.

## 1. Introduction

Reliable intracellular electrophysiological access of neurons is essential to properly measure and interrogate the ionic conductances that underscore neuronal activity. The gold standard technique for intracellular recordings is whole cell patch clamp electrophysiology, which has revolutionized our understanding of the neuronal dynamics of individual excitable cells. In patch clamp electrophysiology, access to the intracellular currents and potentials are obtained at the cost of the health of the cell; most intracellular recordings obtained with patch clamp are extremely labor intensive and provide less than 30-60 minutes of data.[1-3] Instead, evolving technological advances include vertical, nanoscale interface electrodes which are widely utilized to allow intracellular access and recording, with varied structures and interface coupling strategies.[4-9] These techniques can scale intracellular electrophysiological recordings to a large number of cells in extended networks with high spatial resolution. Additionally, due to their nanoscale size, they are minimally destructive and are envisioned to extend the recording duration for days to weeks, all while maintaining the high sensitivity comparable to that of patch clamp recordings. However, most state-of-the-art vertical nanoelectrodes are required to actively deliver electroporation currents to temporarily and destructively permeate neuronal membranes to have intracellular access. Unfortunately, that process is also accompanied with large, irreversible micrometer-scale gas bubble formations[4-6] due to water hydrolysis at the surface of the high-impedance nanoelectrodes.

One successful example of the vertical electrode type is the µm-scale gold mushroom-shaped micro-electrode pillar arrays. The gold mushroom arrays demonstrated intracellular "in-cell" recordings with proper engulfment of the electrodes by the target cell membrane with a short electroporation pulse.[6] Xie *et al.*[7] similarly utilized electroporation with their sub-100 nm, 4x4 platinum nanopillar pad arrays to achieve intracellular action potential recordings that yielded maximum potentials of 11.8 mV post-electroporation, but which eventually attenuated to 30% of the initial amplitude after 2 minutes and to 1% after 10 minutes reaching an extracellular potential of 200 µV. Abbott *et al.*[9] recently attained intracellular recordings by electroporation from arrays composed of 4096 sites built atop complementary metal-oxide semiconductor (CMOS) acquisition electronics. Each electrode contact was composed of nine nanowires coated with low-impedance platinum-black which reduced bubble formation during electroporation. To eliminate the often-uncontrolled destructive process of electroporation, others have used nanowires to mechanically puncture the cell membrane to achieve intracellular recordings. Zhao *et al.*[8, 10] gained intracellular access with their nanowire field-effect transistor into both neuron and cardiomyocytes in culture by probing the cells with an U-shaped 15-nm-diameter Si nanowire. Other nanostructures than 1D nanowires have also been investigated for intracellular recordings. Desbiolles *et al*.[11] · achieved passive intracellular recording with their novel 3D nanovolcano structure comprising of sub-20 nm nanoring. The aforementioned advances have demonstrated intracellular recordings with either cardiomyocyte cells or neurons. Neuronal recording is more challenging because unlike cardiomyocytes, most neurons do not have regular, periodic firing, and also have interspersed axons that result in different waveform and potential dynamics based on the electrode and cell interface locations.[12] Furthermore, neurons possess relatively softer elastic characteristics and lower stiffness that require more contact forces to penetrate their membranes.[13, 14] Of the vertical nanoelectrodes, ultra-sharp nanowires (USNWs) hold the most promise for recording intracellular potentials[15-20] and subthreshold synaptic potentials, and their fabrication techniques afford the scalability needed to achieve simultaneous recordings over extended areas at high spatial resolution.[21]

## 2. Results and Discussion

In this work, we demonstrate a scalable, silicon-based USNW array interface system that natively permeates cell membranes and records intracellular action potentials from cultured rat cortical neurons and from induced pluripotent stem cell (iPSC)-derived cardiovascular progenitor cells (iPSC-CVPCs). Critically, these intracellular recordings are achievable without electroporation. Each contact is composed of a small footprint metallic pad with a diameter of 2 μm, addressing an individual USNW electrode. We observed significantly higher peak-to-peak signal amplitude from recordings with a single USNW compared to those from multiple USNWs, demonstrating the importance of independent electrical addressability for recording high amplitude intracellular potentials.[21] Using sequential focused ion beam (FIB) sectioning to reveal the NW-neuron interface, we assessed the relative position of our metal-coated USNW tips with diameters in the range of 30 – 70 nm with respect to the neuronal cells. Significantly, we detected graded membrane potentials prior to the recorded action potentials from three-dimensional (3D) multi-layered 'tissue' like neuronal networks, justifying the innate ability of our USNWs to record subthreshold potentials, and pharmacologically modulated network activities. Our recordings were not affected by the maturation of the primary culture and glial proliferation, as we achieved high quality recordings from cultured primary rat cortical neurons up to 19 days *in vitro* (DIV). We also observed clear action potential propagation in cardiac networks that can be interrogated by electrical stimulation. Our USNW platform development and their remaining limitations are discussed in detail, uncovering the promise of USNW-neuron interfaces and the challenges set forth to fulfill their full potential.

### 2.1. Fabrication of USNW Arrays

To achieve sub-10 nm vertical USNW arrays, we employed successive and selective oxidation and etching of top-down etched Si USNWs on a Si substrate, invoking standard integrated circuit fabrication technologies in innovative combinations. **Figure 1** and **Figure S1- S6** exhibit details of the fabrication process. Our process involves the dry etching of Si USNW arrays followed by selective oxidation and oxide etching to thin down the NW tips, and the electron-beam lithography and photolithography to attain individually addressable USNW arrays as illustrated in the process schematics (Figure S1). Achieving sub-10 nm diameter USNW tips involves multiple selective oxidations and oxide stripping steps (Figure 1b-f), which yield a fully oxidized Si NW with a smooth surface morphology (Figure 1g). Typical cycles of oxidation and stripping include 10 min – 2 hours of thermal oxidation at 1100 °C and 10 sec – 2 min of buffered oxide etch. To achieve electrical insulation in between individual USNWs, a final oxidation step is used to fully react with the Si NW and the surface of the substrate. The resulting $SiO_2$ USNW is then selectively coated with a Pt metal layer (Figure S3) and a blanket plasma enhanced chemical vapor deposition (PECVD) step is used to deposit 500 nm thick $SiO_2$ passivation layer above the metal leads which is then selectively etched to expose the NW tip (Figure 1b and Figure S4).

Our first set of devices with 300 nm PECVD deposited $SiO_2$ layer was not sufficient to passivate and prevent the delamination of the 10 nm Cr/100 nm Au metal interconnects in *in vitro* cell culture after 14 days (Figure S5a-c). We added a 10 nm thin Ti layer atop the Cr/Au metal leads to promote the adhesion between thinly formed $TiO_2$ layer and the PECVD deposited $SiO_2$ layer on top of the 10 nm Ti layer, and increased the thickness of the $SiO_2$ layer to 500 nm. Accelerated aging experiments by submerging the device into saline solution at 60 °C for three days, equivalent of 15 days at 37 °C, demonstrated robustness of the platform against delamination (Figure S5d, e).

Deposited PECVD oxide layers that had a root-mean-square (rms) surface roughness of less than one nm were too smooth to promote neuronal cell culture adhesion even under the presence of adhesion promoters such as PEI or Matrigel.[22, 23] We calibrated the surface roughness of $SiO_2$ (thermally grown and PECVD deposited, Figure S6), and found that the rough surface with rms roughness of 2-5 nm promotes the cell adhesion according to our initial cell culture experiments. In addition to the $SiO_2$ passivation, we added a 500 nm thick parylene C layer and roughened its surface with $O_2$ plasma treatment (**Figure S7**; rms roughness of 7nm), resulting in better cell viability in comparison to the bare $SiO_2$ surface. The mechanism for this increased cell viability is thought to arise from the inherent nanoscale surface roughness of parylene C and its stable hydrophilic surface for cell adherence after plasma treatment.[24] Devices passivated with additional parylene C coating were used in our experiments. A picture of the overall packaged device is shown in **Figure S8**. The optimized surface materials and roughness were found to help neurite growth (**Figure S12**) and neuronal and cardiac network formation on the USNW array platform (**Figure S13-17**).

Figure 1j shows representative top-view scanning electron microscope (SEM) images of the cultured rat cortical neurons on the surface of our devices. We observed the formation of continuous layers (Figure S14a,b) and of satellite neuronal clusters (Figure S14c,d). FIB sectioning revealed that the cultured neurons exhibit multi-layer structure (Figure 1k and Figure S15: non-colorized SEM image) with extensive neurite connection that suggest excellent cell viability on parylene C-coated vertical USNW arrays. FIB-sectioning in the USNW array region illustrated that the USNW tip resides inside the neuron soma as shown in the time-sequenced SEM images (Figures 1m-o and Figure S16: non-colorized SEM images). While the resolution of these SEM images does not conclusively rule out that the USNW is indeed inside the soma, these images coupled with the recorded and relatively large intracellular potentials support our hypothesis of intracellular access.

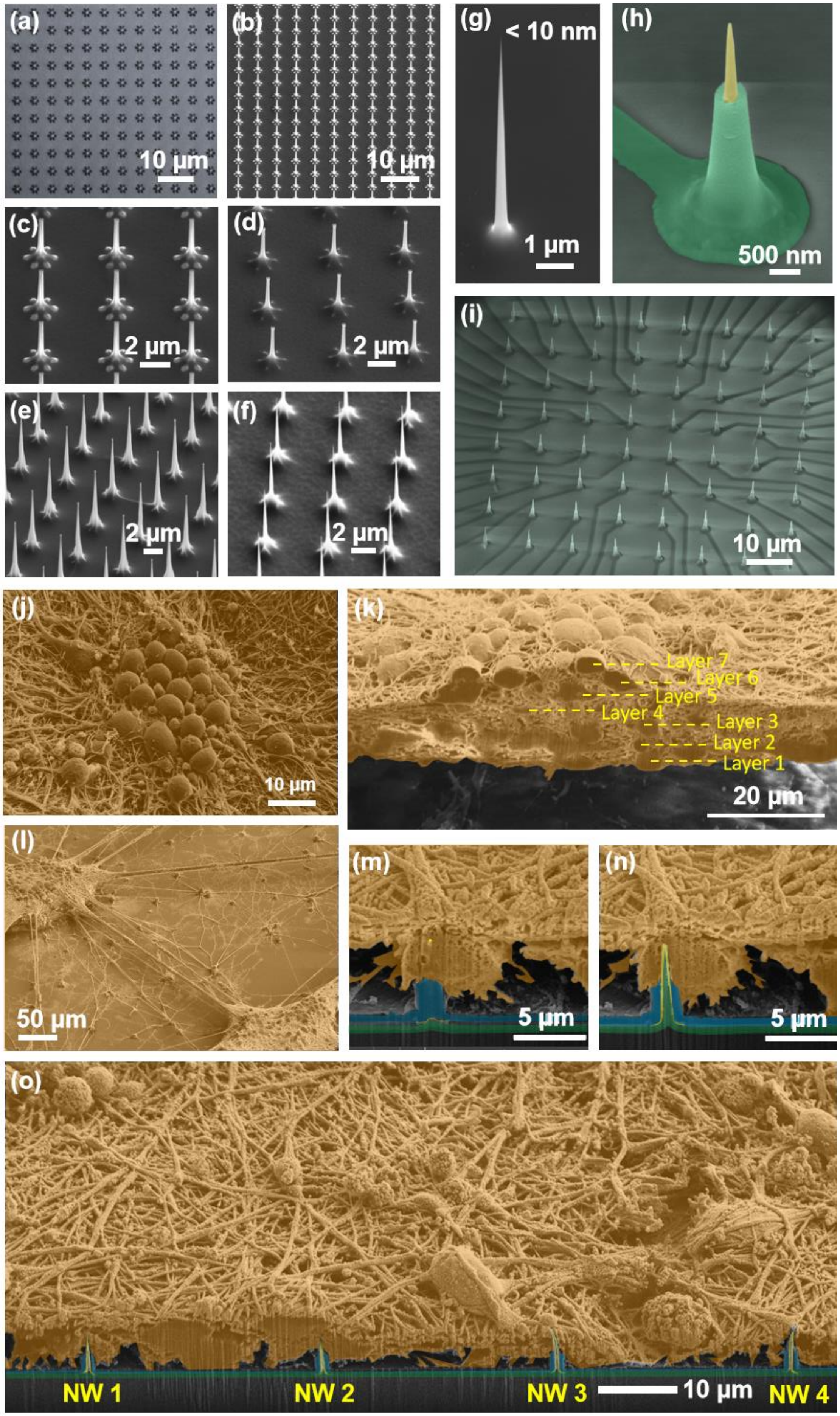

**Figure 1.** Si USNW array fabrication and characterization of structure and the electrode-neural interface**.** Overview of the fabrication process. a) Top-view optical microscope image of patterned Ni dot arrays for Si nanowire etching. The center dot diameter was 800 nm. The peripheral dot diameter was 300 nm. b-d) Example SEM images that shows sequential oxidation and oxide stripping leading to reduced diameter of the NWs in the array, smoothening of the NW surface and the tapered structure. e,f) Example SEM images that shows dry etching, sequential oxidation and oxide stripping leading to smoothening of the surface of the USNWs in the array and the reduction of their diameter to sub-10 nm. g) SEM image of a sub-10 nm Si USNW tip prior to Pt coating. h) SEM image of a single Si USNW showing the tapered structure and exposed Pt tip, and **i,** overall view of Si USNW array. j-o) Colorized SEM images after rat cortical neuron cell fixation showing j) morphology of the cultured rat cortical neurons exhibiting high-density and neurite growth evidencing healthy cell culture and successful network formation, k) cross-section of the cultured neurons exhibiting a multi-layer structure, l) 'satellite'-like interconnected multi-layer structures on the substrate surface, m) sequential FIB sectioning revealing first the tip of the USNW inside the soma and n) the whole USNW/neuron cross-section, o) wide-view SEM image showing the relative position of multiple USNW electrode with respect to the neuron's somas.

As fabricated, the USNWs exhibit impedances exceeding 10 MΩ at 1 kHz (**Figure S9**) that limit a stimulation current to several nanoamperes to avoid electrolyzing water (**Figure S11**a,c). However, in silent or moderately active recording, we don't know *a priori* which USNWs are capable of intracellular stimulation. To enable extracellular current stimulation at the periphery of the USNW array, we selectively coated poly(3, 4-ethylenedioxythiophene):poly(styrenesulfonate) (PEDOT:PSS) on selected USNWs (**Figure S10**) which reduced the electrode impedance by ~ 50X and enabled microampere levels of current to stimulate without causing water hydrolysis (Figure S11b,d). However, for all recordings in this manuscript, we did not use the PEDOT:PSS for stimulation. We stimulated the USNWs for 20 to 40 cycles of 10 nA, 500 µs current pulses injected by a 128-channel Intan RHS2000 stimulation controller and subsequently recorded the electrophysiological activity with the USNWs for several minutes post stimulation (see Supplementary Materials Section 5). Each tested device consisted of four arrays, 32 channels each, with a USNW pitch of 5, 10, 30, and 70 µm (Groups A, B, C, and D, respectively, Figure S4a-d).

### 2.2. *In vitro* Culture Recordings

#### *2.2.1. Recordings from Rat Cortical Neurons*

Multiple *in vitro* cultures from dissociated rat cortical neurons were recorded starting at 7 DIV. Recordings performed from 11 to 19 DIV exhibited diverse characteristics of biphasic and positive monophasic action potentials (**Figure 2**a-h). The raw signals show peak-to-peak amplitudes ranging from approximately 500 µV to 1 mV and upwards towards 10 - 12 mV. The maximum recorded potentials are limited by the Intan amplifier recording range of ±6.4 mV. Using USNW arrays with a different type of recording system, we have previously recorded isolated action potentials with 99 mV amplitude;[21] it is likely that the maximum amplitudes of the action potential reported in this work exceed the ±6.4 mV limits of our recording system. The diverse waveform characteristics on adjacent channels at 11 DIV (Figure 2a) illustrate minimal cross-coupling in between channels in our USNW arrays. We demonstrate representative recordings from all 32 channels in group B (single array with 32 USNWs at 10 µm pitch) from 11 to 19 DIV (Figure 2 a-h). At earlier than 11 DIV (Figure 2a), we only observed a limited number of USNWs with intracellular activity (15/32 channels, determined by the amplitude and waveforms being recorded by each channel over 240 s). Starting from 13 DIV and onwards towards 19 DIV, almost all channels exhibited similar intracellular waveforms (31/32, 32/32, and 31/32 channels respectively for 13, 15, and 19 DIV), indicating synchronous action potential firing and intimate, intracellular USNW-neuron interfaces. Development of multi-layered neural networks is observed directly by gradual increase in spike frequency with the neurons getting mature (Figure 2d,f,h and **Figure S25**b-d) and the evolution of cross-correlation between the channels from low to high synchrony (Figure S25e-h). The synchronized neuronal activity is a common phenomenon in neuronal cultures with multi-electrode arrays[25] and is related to the stage of neuron development and the neuron-glia interactions,[25, 26] and plays a crucial role in complex brain functions and neurological disorders.[27-29] The observed spike synchronicity likely originated from our multi-layered, strictly neuronal cultures with no isolation between neurons by glia and without the *in vivo* molecular heterogeneity composed of extracellular matrices that usually result in heterogeneity in firing in the brain. Clear variance in spike counts across 32 channels in Figure 2g at 19 DIV and the evident variations in graded subthreshold potentials before action potential spikes in Figure 2i confirm non-shorted, strong intracellular USNW-neuron interfaces across the recorded channels. This is further corroborated with evidence presented in this work that shorted nanowires must exhibit small action potential amplitudes (**Figure 4**). By taking the difference in the spike counts (across the recording duration of 240 s) measured from each channel at 13, 15, and 19 DIV to the spike counts of the same channel at 11 DIV, a contour map across the USNW placement is obtained (Figure 2k and **Figure S19**): for longer culture duration, we observed expansion of contour regions from 13 to 19 DIV, signifying an increase in regions of spike activities from few local areas towards the majority of the USNW array. Between different recording days, channels observed shift in USNW-neuron interface conditions (visibly seen for channels 12 and 4 respectively in Figure 2c,g), corresponding to either neuron movement during culture or medium changes between days, which thereby may have exposed the particular USNWs to the culture medium. A representative channel of 7 s recording segment exhibiting large intracellular-like action potentials is shown in Figure 2b,d,f,h across 11 - 19 DIV.

In addition, the electrophysiological activity can be strongly modulated by pharmacological intervention. On 6 DIV, the addition of the $GABA_A$ receptor antagonist picrotoxin (PTX; 33 nM) gradually increased the frequency and the amplitude of action

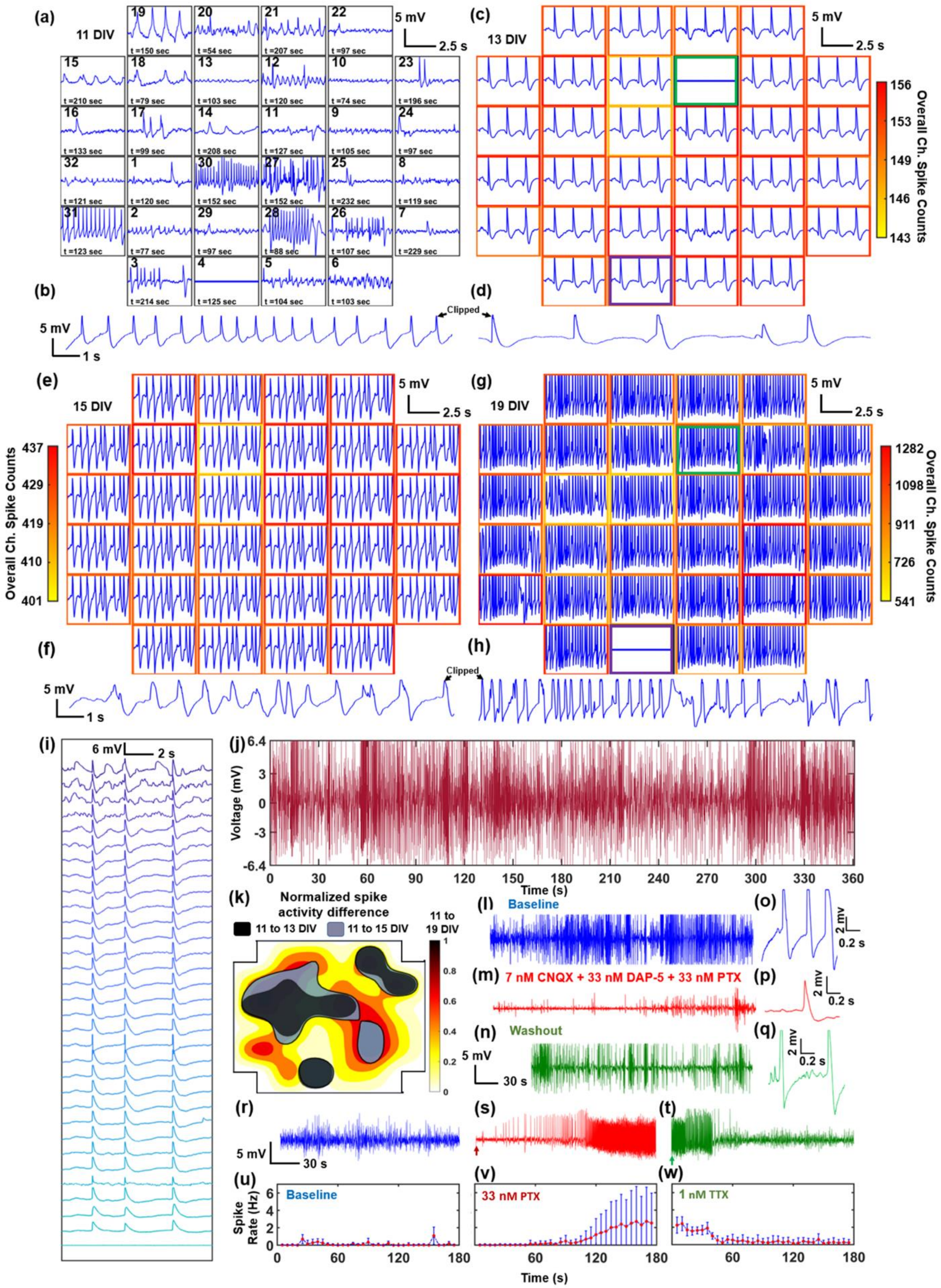
(a)
11 DIV
5 mV
2.5 s
(b)
Clipped
5 mV
1 s
(c)
13 DIV
Overall Ch. Spike Counts
(d)
(e)
15 DIV
(f)
(g)
19 DIV
(h)
(i)
6 mV
2 s
(j)
Voltage (mV)
Time (s)
(k)
Normalized spike activity difference
11 to 13 DIV
11 to 15 DIV
11 to 19 DIV
(l)
Baseline
(m)
7 nM CNQX + 33 nM DAP-5 + 33 nM PTX
(n)
Washout
30 s
(o)
(p)
(q)
(r)
(s)
(t)
(u)
Baseline
Spike Rate (Hz)
(v)
33 nM PTX
(w)
1 nM TTX
Time (s)

**Figure 2.** Example electrophysiological recording from one array (pitch=10 µm) and pharmacological interrogation. a, c, e, g) Overall 5 s electrophysiological recording segments of multiple USNWs from group B (10 µm electrode pitch) at 11, 13, 15, and 19 DIV respectively. b, d, f, h) Close-up image of a 7 s recording segment of selected channel 31 from group B exhibiting intracellular action potential waveforms at 11, 13, 15, and 19 DIV respectively. a) Overall 5 s recording segments at 11 DIV (different time segments are plotted for each channel to illustrate the overall electrophysiological activities at 11 DIV). b) Close-up image of 7 s recording segment, 11 DIV. c) Overall 5 s recording segments at 13 DIV. d) Close-up image of 7 s recording segment, 13 DIV. The green and purple borders highlight silent-to-active and active-to-silent channels 12 and 4, respectively, between 13 and 19 DIV. e) Overall 5 s recording segments at 15 DIV. f) Close-up image of 7 s recording segment, 15 DIV. g) Overall 5 s recording segments at 19 DIV, exhibiting wide variance in spike counts across channels. h) Close-up image of 7 s recording segment, 19 DIV. i) Waterfall plot of action potentials and varying subthreshold potentials arranged in a descending order across 32 channels from group B at 19 DIV. j) Continuous, 6 minutes recording segment from channel 31 USNW from group B (10 µm electrode pitch) at 19 DIV, showing consistent high amplitude without attenuation. k) Contour map of normalized spike activity differences of 32 channels from 13, 15, and 19 DIV with spike activities from corresponding channels at 11 DIV, group B. The black, blue, and red contours represent regions exhibiting spike activity increases from 11 to 13 DIV, 11 to 15 DIV, and 11 to 19 DIV respectively. Contour colormap from 19 DIV is depicted (normalized to the maximum spike activity difference across 32 channels). l, m, n) Sequential recording segment of sequential pharmacological drug test performed at 7 DIV with 7nM CNQX, 33 nM DAP-5, 33 nM PTX, and 1 nM TTX. o, p, q) Representative spike plots from 7 DIV from baseline recording, after CNQX, DAP-5, and PTX application, and final washout. r, s, t) Sequential recording segment of sequential pharmacological drug test performed at 6 DIV with 33 nM PTX and 1 nM TTX. The arrow pointing at the initial section mark the exact moment the drug was applied to the cultured neurons. PTX application results in heightened spike frequencies and TTX application following PTX application and recording results in reduction in spiking activities. u, v, w) Plots of spike rate over time for baseline, PTX, and TTX recording segments. Spike rate change according to pharmacological drug applications.

potentials, consistent with the loss of GABAergic inhibition (Figure 2s). In contrast, we found that the addition of the sodium channel blocker tetrodotoxin (TTX; 1 nM) suppressed firing (Figure 2t). To quantitatively evaluate the effect of pharmacological intervention, we implemented an auto-thresholding algorithm[30-32] on the high-pass filtered recorded waveforms from 32 channels, with a criterion of positive/negative threshold of amplitudes above standard deviation values of the baseline oscillation by 4 – 6.5 σ. As expected, the addition of PTX increased the spike frequency against the baseline recording (Figure 2v) and the application of TTX decreased the spike firing rates (Figure 2w). The change of electrophysiological activity after the drug applications lagged for both PTX and TTX, most likely due to the inherent delay arising from diffusion time of the applied drug molecules through overlaying neuronal layers (Figure 1k) to reach the bottom-most layer of recorded neurons. Additionally, on 7 DIV, another pharmacology modulation study was performed to validate that the pre-spike activities, such as those observed at **Figure S18**c-e, were excitatory postsynaptic potentials (EPSPs). Starting from baseline recording with clear action potential activities (Figure 2l, o), solutions of cyanquixaline (CNQX; 7 nM), D-(-)-2-Amino-5-phosphonopenatanoic acid (D-AP5; 33 nM), and picrotoxin (PTX; 33 nM) were added to respectively block AMPA, NMDA, and $GABA_A$ receptors. Following their application, we observed a clear decrease in action potential and pre-potential activities, as shown in Figure 2m, p and with noticeable reduction in spike frequency as shown in **Figure S21g-i**. After washout, the action potential and pre-spike activities recover (Figure 2n, q), signifying that the pre-spike activities observed prior to drug application were most likely EPSPs.

Overall, we observed more than 46,000 intracellular action potential spikes from the rat cortical neurons across 11 – 19 DIV on the 32 channels of group B. The longest recording segments with continuous, intracellular activities were approximately 6 minutes (Figure 2j and Figure S25). Our measured action potentials retain their amplitudes and shapes over the duration of the recording (Figure 2j and Figure S18). The histogram of the interspike interval (Figure S18c) reveals variability in spike bursting and an exponentially decaying tail, corresponding to action potential refractory periods and spike triggering from random processes, alternating between resting and spiking phases. The mean and mode of the interspike intervals (Figure S18g) range from 500 to 700 ms. For a long recording segment at 11 DIV (Figure S18a), in the first 2 s time snippet (Figure S18b), we observed multiple lower-amplitude spikes within the dampened and broadened temporal response of the USNW array. Such spikes likely correspond to the superposition of high frequency action potential spikes that are faster than the temporal response of the USNW array and in some recordings appeared at a small time offset from the peak of the action potential (e.g., Figure 2h). For this channel, the spike rate was then modulated and was terminated at ~ 210 s. Smaller amplitude potentials that were recorded prior and within the spike trains appear after the 210 s (Figure S18f).

Given that our recording setup is limited to measuring amplitudes below 6.4 mV, we assessed the temporal spread of the waveforms, because it is well established in USNW recordings that longer action potential durations are associated with larger action potential amplitudes. We compared action potential spikes above and below 5 mV, considering that the best amplifier linearity for the Intan amplifier is found was between -5 and 5 mV. Below 5 mV, the temporal spread has a distribution centered around 50 ms whereas those above 5 mV exhibited a temporal spread centered around 75 ms (**Figure S20**a), further suggesting the high sensitivity of our intracellular USNWs. There is a noticeable distribution difference in the spike width, providing evidence the signals were indeed clipped and with recording system with greater limits, the amplitude would have been larger than what was recorded.

### *2.2.1. Recordings from iPSC-derived Cardiovascular Progenitor Cells*

We further investigated the capabilities of our USNW array platform in recording intracellular and network-level activity *in vitro* in iPSC-derived cardiovascular progenitor cells (iPSC-CVPCs)[33] with the culture procedures detailed in the Supplementary Materials Section 3.2. **Figure 3**a shows voltage traces for 52 channels of two separate arrays (out of 64 total channels, 32 channels per array) recorded from the iPSC-CVPCs at day 34 of differentiation (5 DIV on USNW platform). The zoom-in plot of single spikes across the array (Figure 3b) shows a clear peak time delay revealing tissue-wide propagation of action potentials, afforded by the high temporal resolution of the recording (0.033 ms for sampling rates at 30 kS $s^{-1}$). Similar to the electrophysiological recordings from neurons, large intracellular amplitudes from the iPSC-CVPCs were also recorded for the overall duration of the experiment. A representative channel at Figure 3c shows consistent firing with no amplitude decay during a 372 s recording time (beating interval of 2.05 s; Figure 3c,d). The iPSC-CVPC recorded action potentials exhibit clear intracellular attributes: the shape of the spike is a near right-angled triangle with an action potential duration at 50% of the maximum (APD50) of 37.91 ms (Figure 3e) which can be designated as pacemaker iPSC-CVPCs based on their shape. The pacemaker cells have three phases: 0 (depolarization with inward $Ca^{+}$ flow), 3 (repolarization with slow outward $K^{+}$ flow) and 4 (hyperpolarization with slow $Na^{+}$ inward flow).[7, 34] The maximum voltage amplitudes of the action potentials are clipped at 6.4 mV (**Figure S22**), indicating that potentials sensed by the USNW platform are higher than those recorded with our instrumentation.

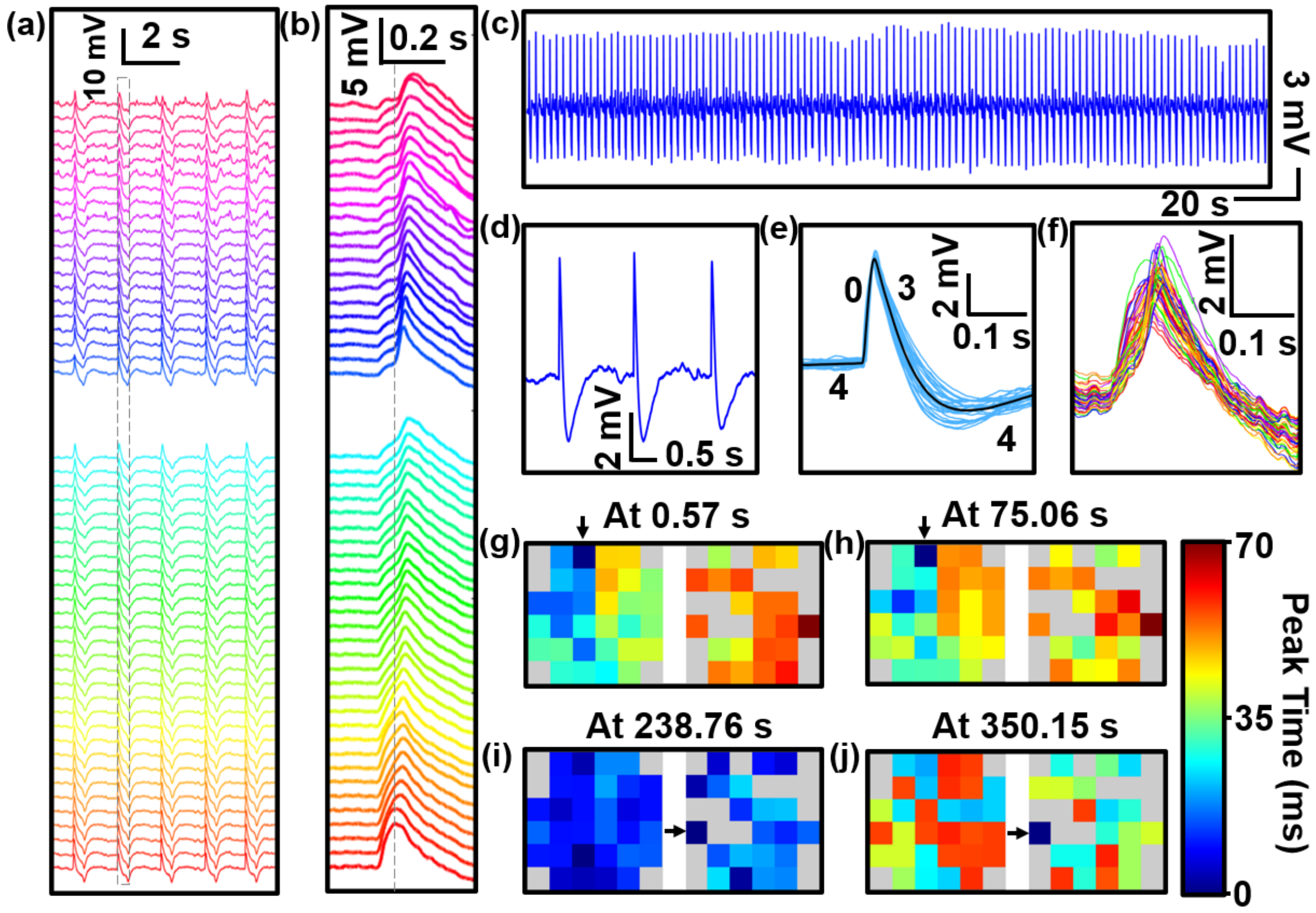

**Figure 3. Network-level intracellular recording of *in vitro* iPSC-derived cardiovascular progenitor cells (iPSC-CVPCs) and active spatiotemporal modulation of action potentials.** a) 52-Channel voltage traces of two arrays recorded from the iPSC-CVPCs at day 34 of differentiation (5 DIV). b) Zoom-in view of the second column of traces in a) shows a single-spiked action potential recording. c) Intracellular recordings of cardiac activity from a representative channel (No. 48) show consistent spiked action potentials with no amplitude decay during the 372 s recording time. d) Zoom-in view of three action potential spikes. e) Intracellular recording of action potentials that is designated as pacemaker based on their shapes. The light-blue traces are 47 randomly selected raw waveforms of a representative spike-sorting action potentials and spike averaged waveform is shown in black. f) Selected time segment of 52-channel voltage traces. g-j) Mapping of action potential propagation patterns across the two arrays at different time points before (at 0.57 s in g) and at 75.06 s in h)) and after (at 238.76 s in i) and at 350.15 s in j)) electrical stimulation. Two intracellular recordings before electrical stimulation (at 0.57 s and at 75.06 s) show action potential propagation from left to right, whereas intracellular recordings after electrical stimulation show an evolution from homogeneous propagation at 238.76 s to reversed direction that originates from right to left, where the action potential propagation direction start from the simulating electrode. The original pacemaker foci location are labeled with arrows. The biphasic-pulse stimulation peak width, amplitude and frequency were 0.5 ms, 10 nA and 1 Hz, respectively.

Our USNW platform enabled us to natively record large intracellular potentials, demonstrating network-level intracellular recording from the tightly connected *in vitro* cultured iPSC-CVPCs 2D tissues (Figure S17) covering two arrays of USNWs. We viewed the difference of the activation time across channels at the same time point (Figure 3f), and we mapped the action potential propagation patterns across the two arrays at different time points before (at 0.57 s and at 75.06 s) and after (at 238.76 s and at 350.15 s) electrical stimulation (Figure 3g-j, and **Figure S23,24**). Two intracellular recordings before electrical stimulation (at 0.57 s and at 75.06 s) show action potential propagation from left to right. We used active electrical stimulation and mapping capabilities to illustrate spatial modulation of the action potential propagation direction within cardiac tissues. Specifically, a USNW–cardiac tissue sample with the original pacemaker foci located at channel No. 32 (Figure 3g) was sequentially paced by stimulator electrodes located at channel No. 52 (see Figure S24c). After electrical stimulation, the activity was synchronized across all channels (Figure 3i, 238.76 s) and then reversed propagation direction from right to left (Figure 3j, at 350.15 s). The new action potential propagation direction starts from the simulated electrode of channel No. 52 (see Figure S24c). The biphasic-pulse stimulation peak width, amplitude and frequency were 0.5 ms, 10 nA and 1 Hz, respectively, applied for 10 s. The results presented here show high-spatiotemporal-resolution electrophysiological mapping and simultaneous interrogation in cardiac tissues for control of cardiac activity, and offer the potential to affect several areas of cardiac research including *in vitro* models for drug-screening, and patient specific models related to cardiac differentiation from progenitor cells or stem cells into damaged tissues with integrated self-mapping and self-modulation functionality.

### 2.3. Performance from Individual Addressability of USNWs

#### *2.3.1. Amplitude Comparison between Numbers of USNWs per Channel*

Finally, we explored the benefit of conducting recordings with individually electrically addressable USNWs. The same fabrication process was implemented for USNW electrodes with 16 and 625 USNWs shorted on a single pad/channel (Figure 4a,c,e). Rat cortical neurons were simultaneously cultured on the three types of samples and they were all recorded from 11 DIV to 19 DIV. Sample recording traces for 10 s are shown for single USNW (Figure 4b), 16 USNWs (Figure 4d), and 625 USNWs (Figure 4f), where we observed the highest recorded action potentials with graded subthreshold potentials from the sample with an individual USNW per channel. The amplitude of the recorded potentials decreased with the increase of the number of USNWs per channel. To quantitatively assess the amplitudes of the recorded action potentials with respect to the number of USNWs per channel, we plotted the histogram of the peak-to-peak amplitudes in a semi-log scale (Figure 4g). We selected channels exhibiting high amplitude signals (15 to 32 out of 32 channels for single USNW, 27 out of 32 channels for 16 USNWs, 30 out of 64 channels for 625 USNWs from 11 to 19 DIV), and compared amplitude distribution with a similar number of detected spikes (2133 spikes, randomly truncated without repetitions from a total of around 6000 detected spikes, from one channel for single USNW, 2133 spikes for 16 USNWs, and 1773 spikes for 625 USNWs). As with pharmacological drug-modulated data analysis, spikes were sorted through an auto-threshold algorithm with bipolar thresholds (Supplementary Materials Section 6). The resulting amplitude distributions for different number of USNWs per site were disparate from one another: as the number of USNWs per site decreased, the distribution of peak-to-peak amplitude increased (Figure 4g). The amplitude distributions were centered around approximately 4 mV and 9 mV, 500 µV and 2 mV, and 60 µV and 300 µV for 1, 16, and 625 NWs per channel, respectively. The different amplitude distributions indicate variability in the capability of each type of USNWs for recording intracellular potentials. However, the plots together with the exhibited potential waveforms clearly demonstrate that individual USNWs per channel yield the highest amplitude action potentials and can record subthreshold potentials. Neither of these attributes can be clearly discerned from the recordings made with 16 and 625 USNWs per channel.

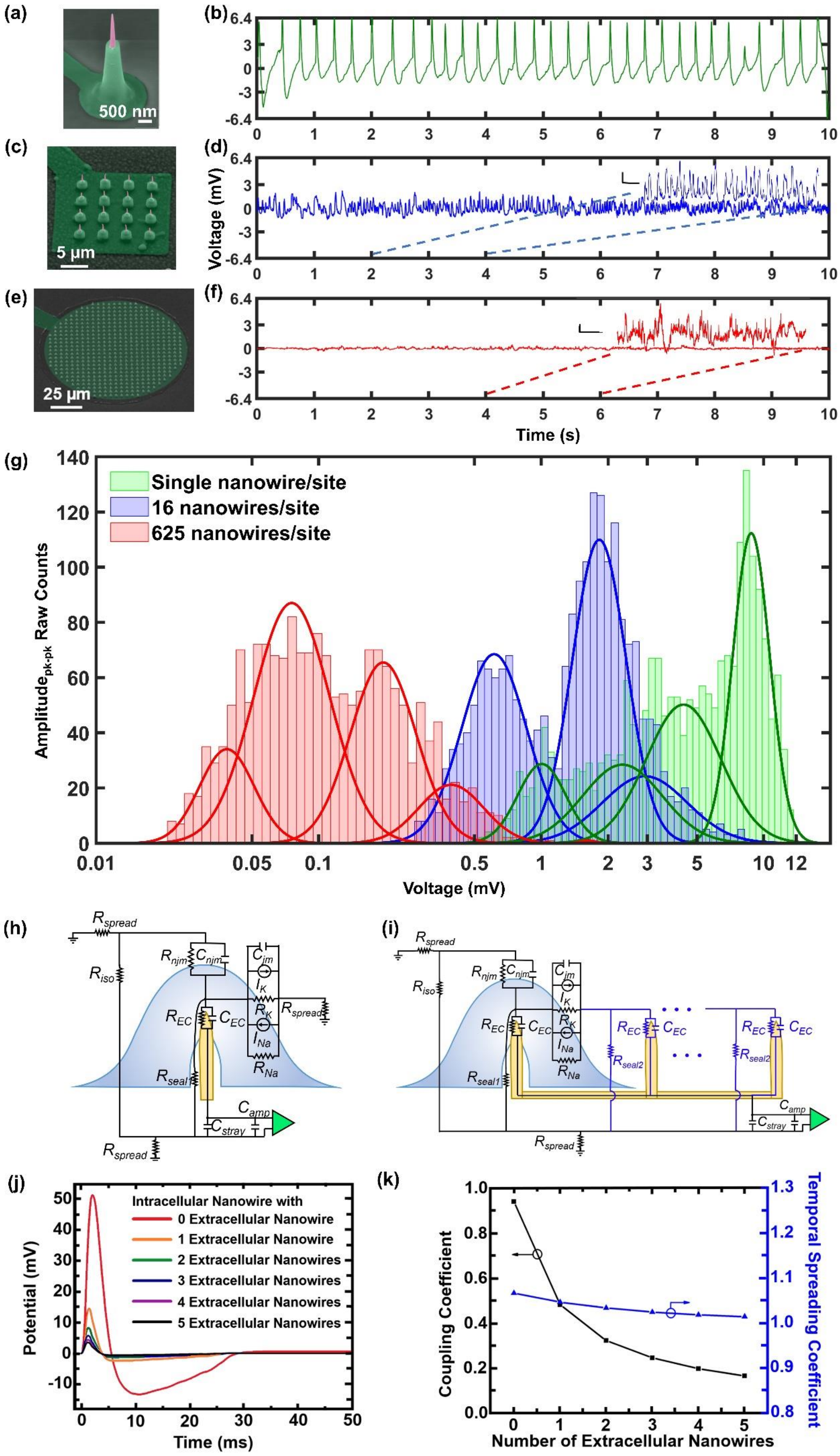

**Figure 4.** Sample recording traces and amplitude contrasts between individually addressable USNW and multiple USNWs per recording sites and small signal circuit model simulation results. a-f) SEM images of individually addressable, single USNW, 16 USNWs, and 625 USNWs and sample 10 s recording trace from recordings performed at 11 DIV. Single USNW - channel 31 from group B (10 µm electrode pitch), 16 USNWs – channel 31 from group A (5 µm electrode pitch), 625 USNWs – channel 21 from group A (5 µm electrode pitch). Vertical scale bars for zoomed-in recording segments for 16 USNWs and 625 USNWs are: 500 µV and 200 µV respectively. The time scale bar for both segments is 250 ms. g) Histogram of peak-to-peak signal amplitude between single USNW and multi-USNWs per site. There are clear, distinguishable differences in the distribution of amplitudes. Two distributions most likely correspond to the intracellular and extracellular recording interface setup the USNW formed with the neurons. h) Circuit model of single USNW penetrating the cultured neuron. i) Circuit model of multi-USNW setup with only one USNW penetrating the cultured neuron. j) Simulated intracellular signal amplitude based on the number of extracellular USNWs connected to the intracellular USNW. k) Simulated plot of coupling coefficient and temporal spreading coefficient versus the number of extracellular USNWs. Signal attenuation decays exponentially with increasing the number of extracellular USNWs.

*2.3.2. Small Signal Circuit Modeling*

The circuit models presented in Figure 4h-i demonstrate a critical design flaw with placing multiple USNWs on a single recording channel, wherein the intracellular potential can be electrically shorted to the grounded extracellular potential. This model builds off the simulations done by Hai *et al.* [35] by investigating the effects of having a portion of the USNWs penetrate a cell body, while the remaining USNWs remain extracellular. The details of this model are discussed in the Supplementary Materials Section 7. Using measurements of transmembrane current during an action potential to ground our models, we simulated the relevant signal paths for several situations. We first modeled the ideal case where a single USNW penetrates the cell membrane and forms a tight seal. We then modeled the effects of adding up to five additional extracellular USNWs on the same recording channel. The resulting potential waveforms shown in Figure 4i demonstrated a maximum signal amplitude for the single USNW case and a decrease in amplitude as we increased the number of extracellular USNWs. Furthermore, when we plotted the ratio of the peak intracellular potential over the peak potential seen by the amplifier (a proxy for signal gain), denoted as the coupling coefficient,[35] we saw that this coefficient decayed exponentially with increasing number of extracellular USNWs (Figure 4k). To develop intuition on the frequency-dependent signal distortion, we computed the ratio of pulse widths of the simulated signal at the input to the amplifier over that of the potential inside of the cell, which we denoted as a temporal spreading coefficient (Figure 4k). These modeling results agree well with the experimental results (Figure 4a-g).

## 3. Conclusion

The present study unequivocally demonstrates the capability of long (> 6 μm) vertical USNW arrays with sharp round tips that are individually electrically addressable for native recording of intracellular potentials. However, the recorded amplitudes on different channels are variable (e.g. Figure 2a), and the source of this variability – likely attributed to the nature of the USNW-neuron interface –was not identified in this study. Potentially, large sample size with more USNW devices and appropriate statistical measures may help quantify the yield and distribution of recorded amplitudes, but it might be a challenging undertaking to conclusively correlate these amplitudes to the experimentally characterized USNW-neuron interfaces. Additionally, there are three main experimental limitations in this study: The first being that our experimental setup requires recording outside the cell culture incubator, where the duration of the recording is limited to several minutes before returning the devices to the incubator in order to maintain constant temperature $CO_2$ concentration and pH levels and to avoid contamination,[36, 37] thereby excluding our ability to assess the longitudinal intracellular recording capacity. This could be tested using cell culture methods compatible with an ex-incubator environment, e.g., HEPES-buffered media and a heated recording platform. Further, sample movements inside and out of the incubator could also alter the USNW-neuron interface to positively or negatively affect the recordings (provide intimate, intracellular interface or expose USNW to culture medium without neuron interface respectively). This limitation can be addressed by performing recordings inside the incubator, which has been demonstrated by Abbott *et al*.[9] The second limitation is that our recording amplifiers are designed for extracellular multi-electrode recordings, which do not perform parasitic capacitance cancellation, and as a result distorts the temporal fidelity of the recordings and broadens the recorded waveforms. This could be addressed using amplifiers with a wide dynamic range for intracellular recordings from thousands of channels, which are available in research labs,[38] where specialized circuits can be custom-designed to restore the temporal resolution of the USNWs for known USNW and parasitic impedances. Lastly, our fabrication method involved multiple dry oxidation cycles to sharpen the USNW tips at 1100 °C, which is not CMOS compatible. We envision acquisition integrated circuits to be integrated on the periphery of our USNW arrays rather than below them to maintain the capabilities and the advantages that this platform brings over the state of the art.

In summary, this work demonstrates that vertical USNW arrays with sub-10 nm tips and individual electrical addressability can record subthreshold and large amplitude action potentials by the natural internalization. We mapped the neuronal/cardiac intracellular activity at the single cell resolution, observed clear synaptic network activity between neurons, and cardiac activity propagation in the extended networks, and manipulated the cardiac signal propagation direction by electrical stimulation. Using both experiments and simulations, we validated that multiple USNWs per single channel reduce the amplitude and sensitivity of the recordings compared to single USNW per channel recording. We believe that these results underscore a significant advancement in our understanding and control over the USNW-neural interface. Novel integration of the USNW with depth probes or 2D soft substrate may be applied *in vivo* for the potential intracellular recordings from intact brains. With the demonstrated sensitivity, this platform paves the way for novel scientific and technological undertakings that aim to establish large-scale bidirectional biotic-abiotic interfaces with intracellular access for drug screening, disease modeling, and beyond.

## 4. Experimental Section/Methods

*USNW Electrode Design, Fabrication, Packaging, and Characterization*:

Our Si USNW platform consists of total of 128 channels divided into 4 subgroups with different USNW pitch of 5, 10, 30, 70 μm (Figure S4) to provide a judicious range of USNW density to maximize the probability of membrane permeability by changing the electrode-neuron interface tension. Generally, narrower pitch requires taller nanoelectrodes for effective membrane penetration, which is accounted for with our USNW's height, standing between 6 to 7 μm.[39]

The array fabrication begins with e-beam lithography (EBL) patterning of center dots with 800 nm diameter and peripheral dots with 300 nm diameter in the resist on the prime-grade p-type doped Si substrate (Figure 1a and Figure S1,2). 200 nm of Ni is then evaporated and lifted off, forming dot-like arrays that form the etching masks for Si. A $SF_6/C_4F_8$ based inductively coupled plasma (ICP)/reactive ion etching (RIE) process is then utilized to selectively etch the uncovered Si substrate by 6−7 μm to form vertically standing Si nanowires (Figure S1,2). These nanowires initially have flat tips following the planar Ni discs. The resulting nanowires are then processed through multiple thermal oxidations and buffered oxide etch (BOE) cycles to provide smooth surface, achieve desired tapering for robust support, and thin USNW tips as shown in Figure S1,2. Following the USNW formation, the whole device undergoes a long, unmasked, thermal oxidation to fully oxidize the vertically standing USNWs and the surface of the substrate to electrically isolate all USNWs. Subsequently, patterns of USNW electrode and center interconnects are aligned and defined by EBL, and 10 nm/100 nm of Cr/Pt are deposited by conformal electron-beam evaporation through sample rotation to provide independent electrical addressability for each USNW (Figure S3). To avoid electrochemical coupling and corrosion in the culture medium, the metal interconnects were passivated by using PECVD $SiO_2$. Control experiments demonstrated that a 500 nm thick $SiO_2$ passivation layer is needed for stable passivation against delamination in a wet environment (Figure S5). Additionally, we employed a 500 nm thick parylene C layer for both passivation and improvement of adhesion of cells and substrate. Both the parylene C and the $SiO_2$ layer were selectively etched from the USNW tips by recessing a resist layer and employing an $O_2$ plasma etch for parylene C and BOE etch for $SiO_2$, exposing the Pt coated USNW tips while the remained of the USNW remains unaltered (Figure S6,7). Alongside $SiO_2$, parylene C exhibits excellent mammalian cell adhesion properties after proper surface preparation via $O_2$ plasma treatment.[24] For *in vitro* cell culture, the final device is bonded with a culture chamber via polydimethylsiloxane (PDMS) application to delineate the region of cell culture interest with the rest of the device. Finally, the USNW electrodes are electrically connected with flexible flat cable (FFC) via anisotropic conductive film (ACF) bonding (Figure S8). The resulting USNW structure is composed of an exposed metal tip and a passivated bottom layer for good electrical isolation. The ideal USNW tips are sharpened to sub-10 nm range before conformal metal coating with proper masking and optimized etching process (Figure 1i, Figure S1f). Overall, USNWs formed consistently throughout the array with uniform electrochemical impedance. A representative characterization result of the 128 channel devices showed 100% yield and average impedance of 14.95 MΩ at 1 kHz in Figure S9.

*Rat Cortical Neuron Culture on USNW Array Electrodes*:

The integration of USNW array electrodes with cells followed immediately from fabrication and packaging of the devices, which involved the following steps (steps 1 to 4 were also applied before iPSC-CVPCs cultures):

(1) The USNW array device was sterilized by 1) DI water rinse for three times, and 2) 70% ethanol for at least 30 min.

(2) The device was washed with phosphate buffered saline (PBS) for three times followed by placing 20 µl drop of (0.1% w/v) Polyethylenimine (PEI, Sigma-Aldrich) solution on the USNW array and incubating them at 37 °C incubator for 1 hour.

(3) The PEI was aspirated and washed with 500 µl double distilled water (dd$H_2O$) for 4 times. The device was dried in the incubator for 5 hours/overnight.

(4) A spot of 20 µl laminin (20 µg ml$^{-1}$, Sigma-Aldrich) was added to the USNW array and incubated at 37 °C for 1 hour.

(5) Neuronal cell culture medium (Neurobasal (Thermo Fisher Scientific) +2% B27 (Thermo Fisher Scientific) +1% P/S (Corning) + 10% FBS) was prepared for plating rat cortical neurons on the USNW arrays.

(6) The cryopreserved rat neurons cryovial (from Thermo Fisher Scientific) was removed from the liquid nitrogen storage container, warmed in a 37 ℃ water bath for exactly 2.5 min, sprayed the outside with 70% ethanol, wiped dry, and placed in a tissue culture hood.

(7) The contents of the cryovial were carefully transferred to a 15 mL centrifuge tube using a 1 mL pipettor. The inside of the cryovial was carefully washed with 1 mL of room temperature neuronal cell culture medium (~1 drop $s^{-1}$). 3 mL of room temperature neuronal cell culture medium was slowly added to the tube (~1-2 drops $s^{-1}$). The contents were carefully mixed by inverting the tube 2-3 times. The total number of cells in suspension was determined via hemocytometer count.

(8) The cells were concentrated by centrifuging 1100 rpm for 3 min. Laminin was aspirated and 20 µl cell suspension was plated directly on the USNW arrays with the cell density at 200k cells/array. The USNW devices with seeded neurons were incubated in a cell culture incubator at 37 ℃, 5% $CO_2$ for 40-50 min. Next, 500 µL of warm Neurobasal medium was carefully added to chamber from the side of the device. To avoid adding the medium too fast to cause the detachment of the adhered neurons on the USNWs, the USNW devices with seeded neurons were put back to the cell culture incubator at 37 ℃, 5% $CO_2$ for another 35-40 min. Then, another 500 µL of warm Neurobasal medium was carefully added to chamber from the side of the device to reach a volume of 1.5 ml per device. Finally, the USNW devices with seeded neurons were put back to the cell culture incubator at 37 ℃, 5% $CO_2$.

(9) On the next day, the medium was changed to Brainphys Complete (Brainphys basal medium (Stemcell technologies)+1% N2 supplement (Thermofisher) +2% B27 supplement (Thermofisher) +1% Pen/Strep (Gibco/Life Technologies)) with the addition of the following supplements: Brainderived Neurotrophic Factor (BDNF, 20 ng ml$^{-1}$; Peprotech,), Glia-derived Neurotrophic Factors (GDNF, 20 ng ml$^{-1}$; Peprotech), ascorbic acid (AA, 200 nM; Sigma), dibutyryl cyclic AMP (cAMP, 1mM Sigma) and laminin (1 μg ml$^{-1}$; Invitrogen). Then, half the medium was replaced with fresh medium every other day with Brainphys Complete and supplements till 7 DIV, then switched back to Brainphys Complete with no added supplements.

*FIB-SEM Characterization of USNW-Neuron Interface*:

To precisely investigate the biological interface formed between the USNWs and the cultured neurons after the completion of electrophysiological recording sessions, a sequential FIB cut was performed to reveal the cross-section of USNW regions after appropriate Pt plating for protection from ion-milling damages. The successive FIB cuts show both proper engulfment and clear intracellular penetration of the USNW into the cultured rat neurons' soma, which align well to the positive-phase potentials measured in our electrophysiological recordings, shown in followed analysis. The intracellular nanoelectrode-neuron interface observed here is based on native penetration deriving from the ultra-sharp USNW tips, without any additional surface chemical treatments such as peptide-modification[40] and widespread electroporation for increased cell membrane permeability. The $SiO_2$ USNWs were rigid and free-standing for stable interaction with the neurons without any USNW breakage.

*Neuronal Recording and Pharmacological Drug Application*:

Three minutes of baseline activity was recorded, followed by 33 nM application of the $GABA_A$-R blocker picrotoxin (PTX, Tocris) to the device chamber. The activity was then recorded for 3 to 5 minutes to observe the effect of blocking inhibition on the electrical activity of the neuronal network. Finally, 1 nM tetrodotoxin (TTX, Abcam) was added to the solution to block voltage-gated sodium channels ($Na_v$) and prevent generation of action potentials.

**Supporting Information**

Supporting information is available in the online version of the paper from the Wiley Online Library. Reprints and permissions information is available online. Correspondence and requests for materials should be addressed to S.A.D.

**Acknowledgements**

This work was performed with the gracious support of National Science Foundation Award No. 1728497 under the stewardship of Dr. Khershed Cooper and of the National Institutes of Health Award No. NBIB DP2-EB029757 under the stewardship of Dr. Michael Wolfson. S.A.D., J.S.M. and R.L. also acknowledge the gracious support of the UC-National Laboratory in Residence Graduate Fellowships (UC-NLGF), Award No. 477131. This work was performed, in part, at the Center for Integrated Nanotechnologies, an Office of Science User Facility operated for the U.S. Department of Energy (DOE) Office of Science. Los Alamos National Laboratory, an affirmative action equal opportunity employer, is managed by Triad National Security, LLC for the U.S. Department of Energy's NNSA, under contract 89233218CNA000001 and Sandia National Laboratories (Contract No. DE-AC04-94AL85000) through a CINT user proposal. This work was performed in part at the San Diego Nanotechnology Infrastructure (SDNI) of UCSD, a member of the National Nanotechnology Coordinated Infrastructure, which is supported by the National Science Foundation (Grant ECCS-1542148). The authors thank technical support from the Integration Laboratory at CINT and from the nano3 clean room facilities at UC San Diego's Qualcomm Institute. R.L. and S.A.D. would like to acknowledge inspiring technical discussions with Dr. Renjie Chen, and Dr. Atsunori Tanaka, as well as discussions and technical support from Dr. Katherine L. Jungjohann, Anthony R. James, Douglas V. Pete, and Denise B. Webb of Sandia National Laboratories.

S.A.D. conceived and led all aspects of the project. R.L. developed the fabrication process with S.A.D. R.L. fabricated the platform with input and training from J.N., J.Y. M.E.P. and J.S.M; Y.G.R. and L.A.H. participated in the fabrication; R.L. performed the FIB sectioning and SEM imaging. R.L. and R.V. performed electrochemical characterization; R.L. and S.H.L. performed electrophysiological recordings; D.P. and G.R cultured the primary rodent neurons under supervision of A.G.B.; A.D.C. cultured iPSC-derived cardiovascular progenitor cells under the supervision of K.A.F.; A.M.B. and S.A.D. carried out the electrochemical interface modeling. J.L., Y.T., R.L., K.J.T. and S.A.D. analyzed the data and wrote the manuscript. All authors discussed the results and contributed to the manuscript writing.

Ren Liu, Jihwan Lee, and Youngbin Tchoe contributed equally to this work.

# Supporting Information

**Ultra-Sharp Nanowire Arrays Natively Permeate, Record, and Stimulate Intracellular Activity in Neuronal and Cardiac Networks**

*Ren Liu, Jihwan Lee, Youngbin Tchoe, Deborah Pre, Andrew M. Bourhis, Agnieszka D'Antonio-Chronowska, Gaelle Robin, Sang Heon Lee, Yun Goo Ro, Ritwik Vatsyayan, Karen J. Tonsfeldt, Lorraine A. Hossain, M. Lisa Phipps, Jinkyoung Yoo, John Nogan, Jennifer S. Martinez, Kelly A. Frazer, Anne G. Bang, Shadi A. Dayeh**

**This file includes Supplementary Methods, Supplementary Figures, and Supplementary References:**

**1. Fabrication, optimization process and packaging of individually electrically addressable Si USNW array devices**

1.1 RIE/ICP dry etching and thermal oxidation processes for Si USNW etching, thinning, and sharpening.

We performed mechanistic studies to obtain tapered USNWs with a larger base diameter, since the inverted cone structure provides the USNWs enough mechanical strength and stabilizes the USNWs' shape and potentially the electrical interface with neurons. The overview of the key fabrication process steps is shown in the schematic figures in Figure S1a-h, with cross-sectional overview of the fabricated USNW shown in Figure S1i. Initially, dot patterns (center dot diameter was 800 nm and peripheral dot diameter was 300 nm) and alignment markers were patterned on the Si substrate by using JEOL EBL (JBX-6300FS), and Ti/Ni (10 nm/ 200 nm) was deposited by using electron-beam evaporation (Figure S1j). After lift-off in acetone for 2 hours, samples were cleaned by IPA and $N_2$ blow dry. Next, USNWs and markers were etched by using Plasmatherm SLR-770 DRIE / ICP Etcher (Figure S1k). The reactive-ion etching (RIE) power was 10 W, the inductively coupled plasma (ICP) power was 700 W, the chamber pressure was 23 mTorr, the electrode temperature was 20 ℃, the gas flow of $SF_6$, $C_4F_8$, Ar were 20 sccm, 40 sccm and 40 sccm, respectively (etch rate is 110 nm $min^{-1}$). Then, we used diluted nitric acid ($HNO_3$:$H_2O$ = 1:10) to etch the Ni etching mask residue, followed by cleaning polymer residue from the Si NW etch with Piranha solution (sulfuric acid 30%:hydrogen peroxide solution = 3:1), rinsing with deionized water, drying with $N_2$ gas (Figure S1l). Next, the sample was placed on a quartz boat and loaded into a thermal oxidation furnace tube. The thermal oxidation temperature was 1100 ℃. A layer of $SiO_2$ was formed at the surfaces of the NW and of the substrate after the thermal oxidation step. We used wet etching to remove the $SiO_2$ by using buffered oxide etch (BOE: 6:1 volume ratio of 40% $NH_4F$ in water to 49% hydrofluoric acid (HF) in water). To avoid the whole

NW being thermally oxidized, we used dry thermal oxidation rather than the wet oxidation. Since the wet oxidation rate is more than 10 times faster than the dry oxidation rate, dry oxidation is more controllable for thinning small USNWs, while wet thermal oxidation may be suitable for thick microwire thermal oxide growth and the creation of a smooth interface at the oxide/Si interface. We relied on measurements of the NW diameter after each oxidation and oxide stripping step with the SEM because deviations from the thermal oxidation rate calculated with Massoud Model[1] were observed and as expected due to slow oxidation in NWs. We practically repeated the thermal oxidation and wet etching for a few times to get the USNW as sharp as a few nanometers of tip diameter (Figure S1m-o). The process is repeatable and shows 100% yield (visual inspection of the USNWs' tip condition via SEM) of all samples as two examples shown in Figure S2a-h.

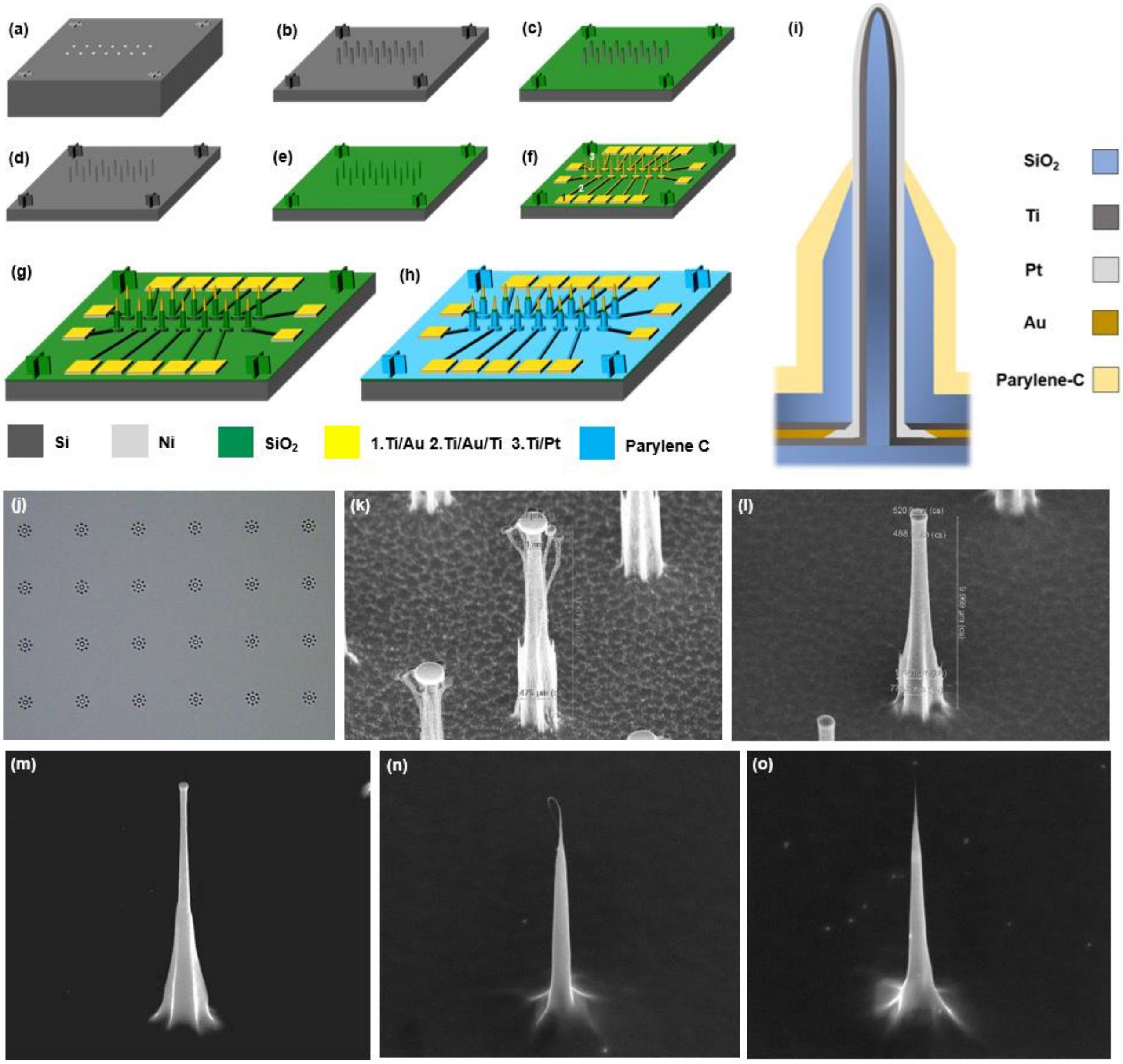

**Figure S1.** Overview of the fabrication process with a) Patterning Ni dots on Si substrate**,** b) Dry etching micro/nanowires, c) Thermal oxidation, d) Wet etching of oxidation layer, e) Thermal oxidation, f) Definition of the metal leads and contacts by electron-beam and photolithography, and g) 500 nm PECVD $SiO_2$ passivation h) 500 nm parylene C passivation and selective removal at the NW tip. i) Cross-section overview of the fabricated USNW. j) Top-view optical microscope (OM) image of patterned Ni dot arrays for Si NW etching. The center dot diameter was 800 nm. The peripheral dot diameter was 300 nm. k) SEM image of as-etched Si NWs showing thin peripheral NWs, mostly collapsed and a central NW with narrower tip than its base. l) After Ni mask etching, the NW was subjected to the first oxidation and BOE etching. The tip diameter was 200 nm and the NW length was ~10 μm. m-o) Sequential oxidation and oxide stripping leading to (1) smoothening of the NW surface, (2) reduction of its diameter to sub-10 nm, and (3) reduction of its height to ~ 6.5 μm. The final NW is fully oxidized.

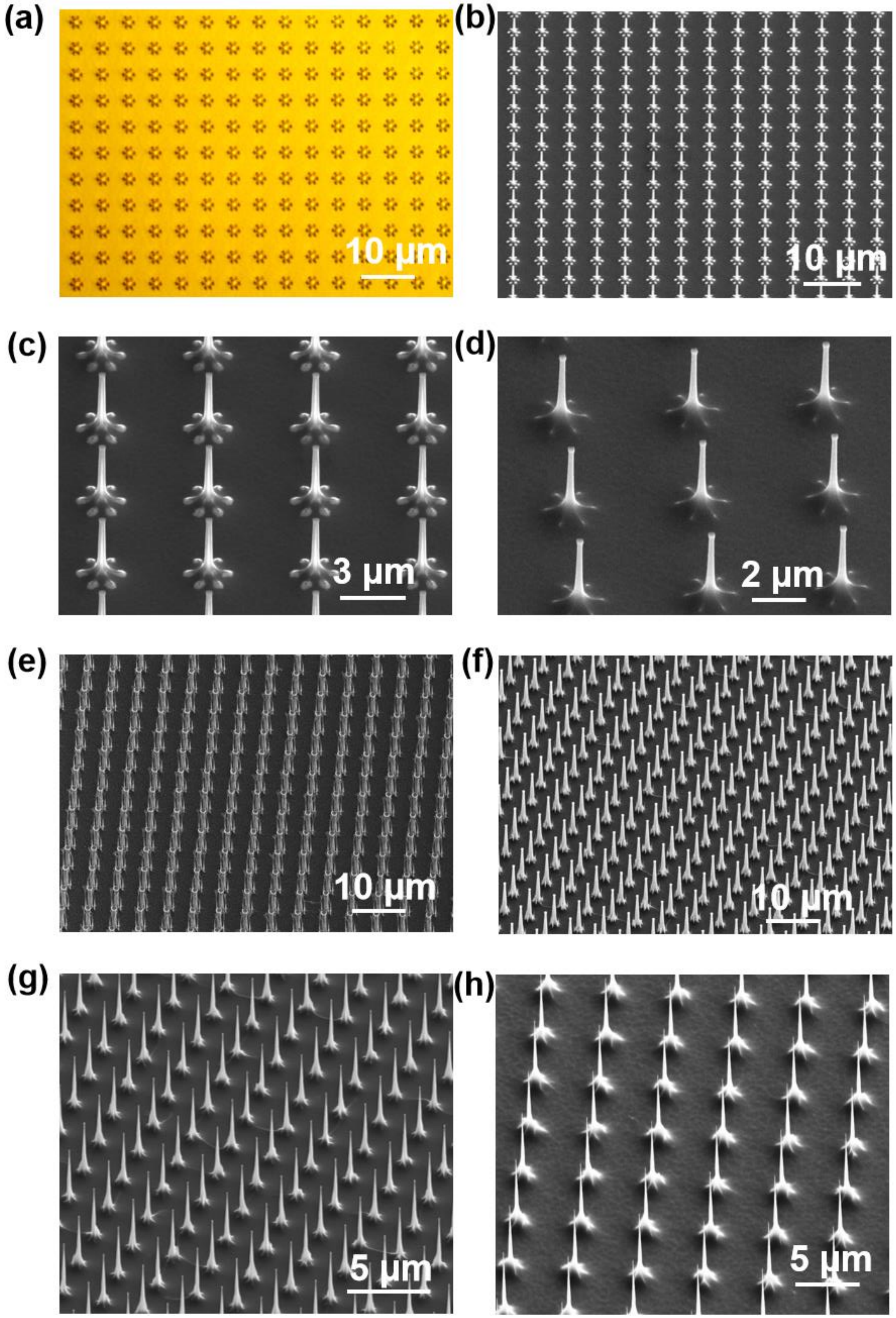

**Figure S2.** a) Top-view OM image of patterned Ni dot arrays for Si NW etching. The center dot diameter was 800 nm. The peripheral dot diameter was 300 nm. b-d) SEM images of an example that shows sequential oxidation and oxide stripping leading to reduced diameter of the NWs in the array, smoothening of the NW surface and the tapered structure. e-h) SEM images of another example that shows dry etching, sequential oxidation and oxide stripping leading to smoothening of the surface of the NWs in the array and the reduction of their diameter to sub-10 nm.

1.2 Fabrication for electrodes and metal interconnects

We performed a final step of 400 nm thermal oxide growth on the surface after the NW tip reaches a few nanometers. After the sample cleaning by acetone airbrush, rinsing by isopropyl alcohol (IPA), drying by $N_2$, dehydration at 180 °C on a hot plate, $O_2$ plasma at 200 W for 1 min, we spun coated a layer of electron beam resist (950 PMMA A6) with spin-coating condition of 4000 rpm, 500 rpm $s^{-1}$ and 40 s, and baked the sample at 170 °C for 10 min. Then, we used EBL to write the pattern with the dosage of 1800 μC $cm^{-2}$ for metal leads and 3000 μC $cm^{-2}$ for dots (aligned with the USNW) under the electron beam condition at 100 kV and 6 nA, followed by development (methyl isobutyl ketone (MIBK):IPA = 1:3) at 0 °C for 2 min, which opens the electron beam resist (950 PMMA A6) on the surface of the USNWs to connect to the metal leads. We used $O_2$ plasma for removal of 3-5 nm-thick PMMA residue on the pattern at 100 W for 1 min. A conductive layer, e.g. 10 nm Ti / 100 nm Au or Pt, etc., was uniformly coated on the sidewall of the USNW to form a conductive layer and connect the electrode tip to the metal leads (Figure S3a,d,e). Then, photolithography was used to pattern the outer large electrodes and interface/probing pads (Figure S3b). Finally, we used plasma enhanced chemical vapor deposition (PECVD) to deposit a layer of 300 nm $SiO_2$ at 350 °C (for first experiment) on the surface of the device with a hard mask (Si wafer with a pocket opening in the center) to avoid the $SiO_2$ deposition on the around the interface/probing pad (Figure S3c). Then, we spin-coated a layer of electron beam resist (PMMA 950 A4). After 10 min of hard bake at 170 °C, 10 – 20 s of $O_2$ plasma with 200 W was applied to clean the thin PMMA layer on the tip of the USNW. Then, we used BOE to etch the $SiO_2$ on the tip. After cleaning PMMA by acetone and $O_2$ plasma, the USNW has ultra-sharp tip (~30 nm) is shown in (Figure S3f). The sample with 128 channels of individually addressable USNWs are distributed for 4 arrays with different USNW pitch as 5 µm, 10 µm, 30 µm, 70 µm as shown in the Figure S4a-d, respectively. The process also enables variable USNW height as an example shown the USNW electrode with height of 2.9 µm to 9 µm in Figure S4e-i.

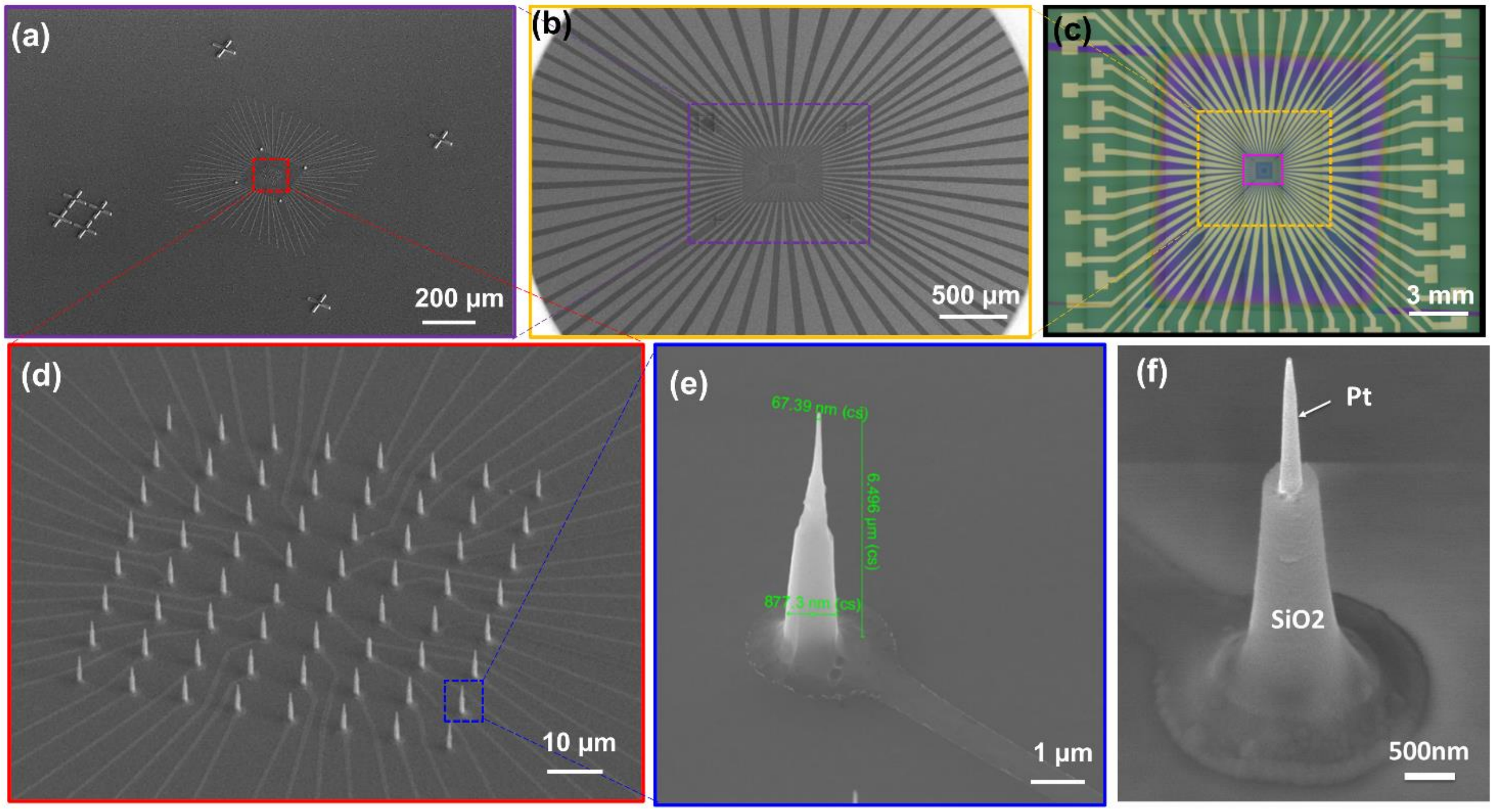

**Figure S3.** Metallization and Passivation of USNW and Electrode. a) SEM image of the e-beam lithography patterned portion of the USNW array, where metal lead extensions were contained within the e-beam lithography patterned alignment marks within the e-beam lithography writing field. b) SEM image of the device after the photolithography patterning and deposition of the extension metal leads to the interface/probing pads. c) Optical image of the device with passivation surface of $SiO_2$ deposited by PECVD. d) Zoom-in images of the USNW array in a). e) SEM image of a single USNW coated with a thin (30 nm) Pt layer conformally deposited by e-beam evaporation. f) USNW with passivation of its base by PECVD deposited $SiO_2$.

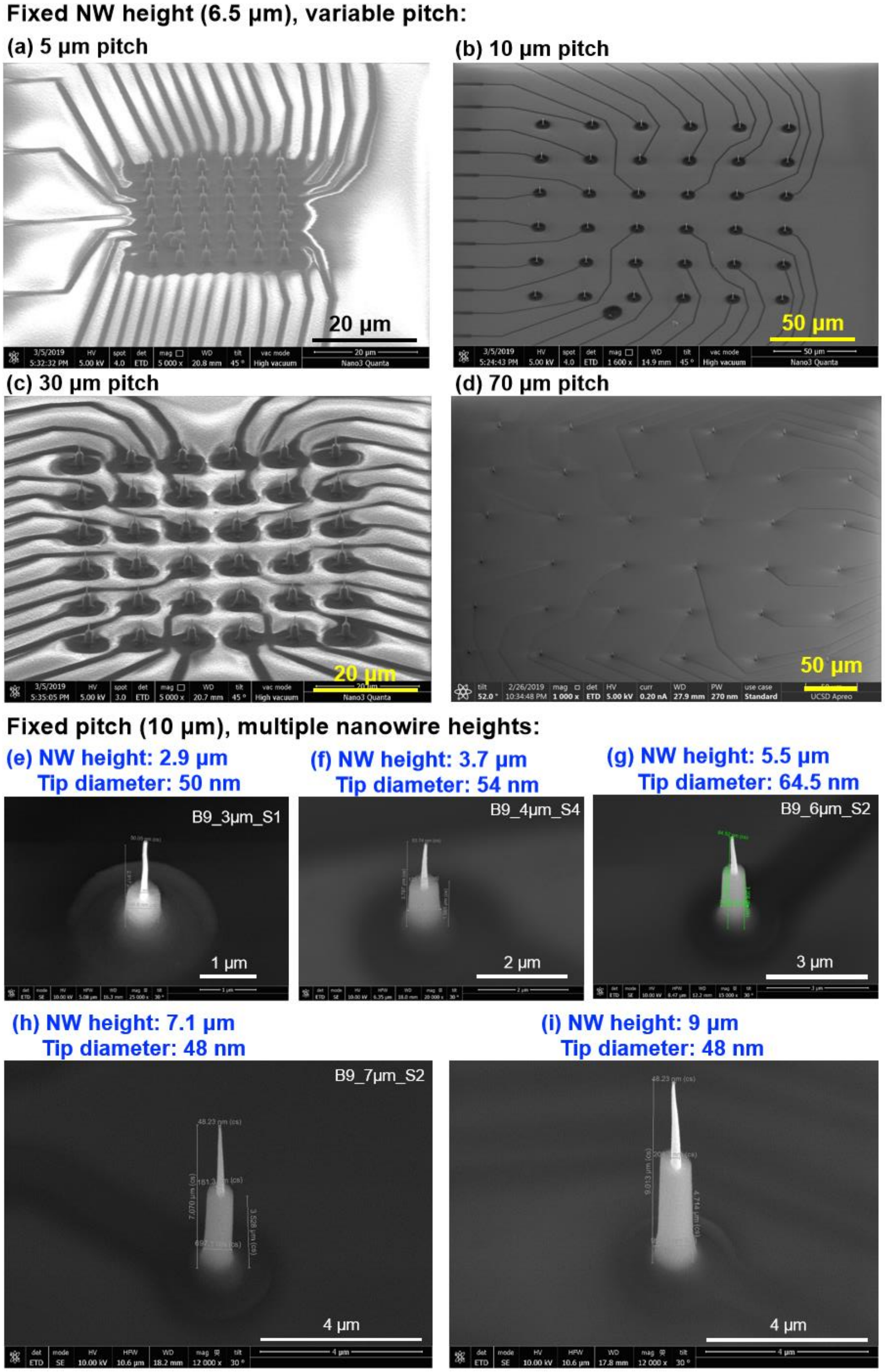

**Figure S4.** a-d) SEM image of the vertical USNW device arrays. Each device consisted of 128 USNWs in total, and 4 arrays, each with 32 USNWs and different pitch as labeled in a-d) (groups A to D). The 4 USNWs at each corner are not connected to the metal leads. **e-i)** The process also enables fixed pitch and variable USNW height as shown in panels e-i.

1.3 Optimization of passivation

After culturing the cells on the devices for 14 days, a layer of 300 nm thick $SiO_2$ on Cr/Au metal leads peeled off as shown in Figure S5a. A FIB cross-section of the peeled-off region was cut and showed a gap between the $SiO_2$ and the metal lead (Au or Pt) (Figure S5b), because of the poor adhesion between $SiO_2$ with the noble metals. Another problem was that the leads patterned by electron beam lithography (EBL) had tall edge beads (Figure S5c) because the conformally electron beam deposited metal caused difficulty of the lift-off at the edge of the pattern. To solve the peel-off problem, we deposited an adhesion layer of metal (10 nm Ti) on top of the previous Au or Pt, which enhanced the adhesion between passivation and the metal layers. Then, during EBL of the center leads, we also changed the electron beam resist from a single layer of e-beam resist (PMMA 950 A6) to double layers of resists (Methyl methacrylate (MMA) (8.5) A6 and PMMA 950 A6). The usage for MMA/PMMA bilayer in EBL helps lift-off of metallic structures, since MMA/PMMA bilayer gives an undercut resist profile to avoid metal coating on the sidewall of the resist during the electron beam deposition. We also increased the thickness of the $SiO_2$ passivation layer to 500 nm. After these steps of modification, we did an accelerated aging test of the devices by submerging them in saline solution at 60 °C for 3 days, which is equivalent to the condition of 37 °C for 15 days (cell culture temperature and timeline). The devices showed clear surface (Figure S5d) after the aging test. Two cross-sections of FIB-SEM images demonstrated good adhesion between metal and passivation layer (Figure S5e) and no edge beads (Figure S5f).

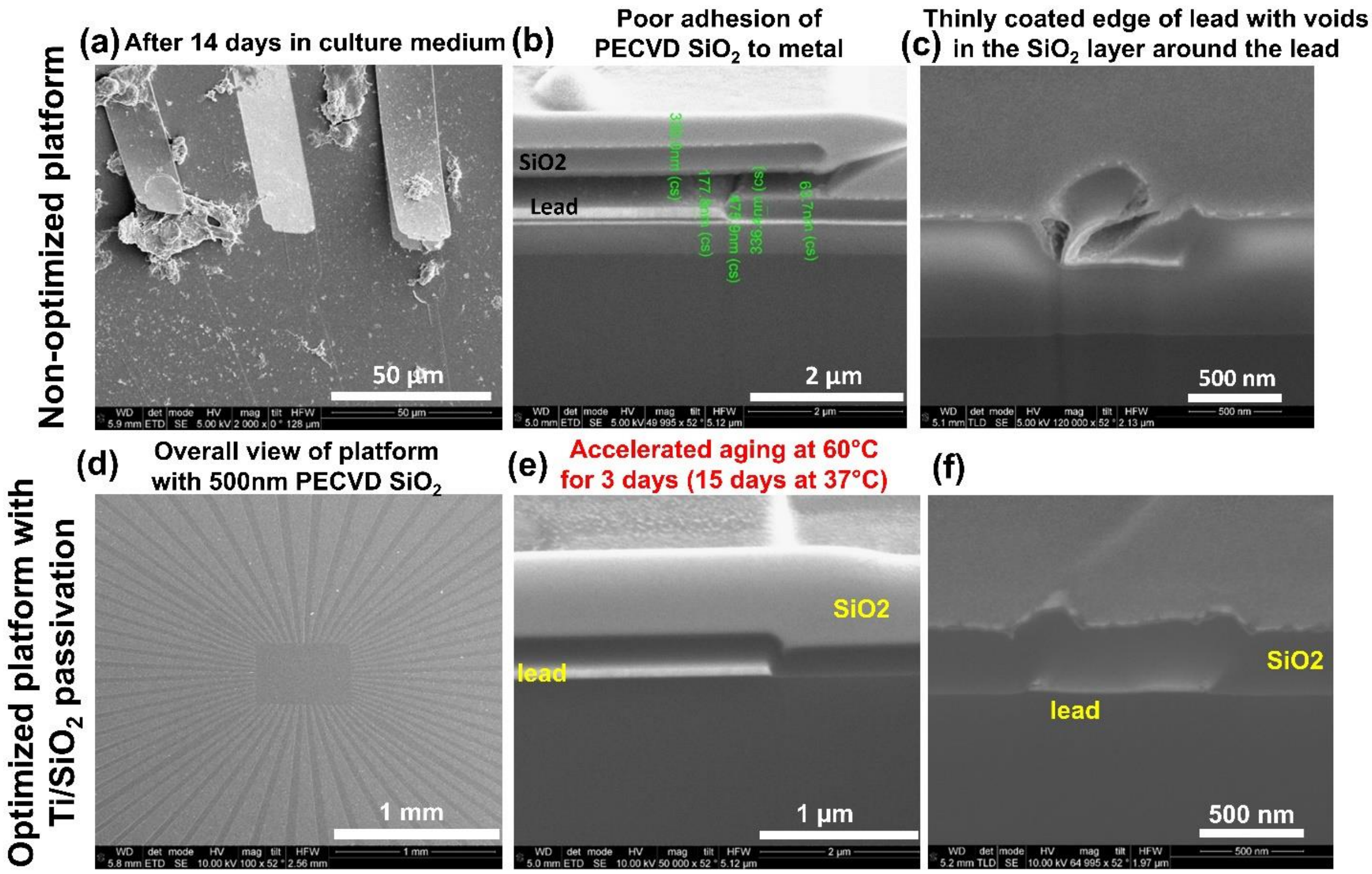

**Figure S5.** a-c) 300 nm PECVD $SiO_2$ passivation layer on top Cr/Au metal leads show gross delamination in panel (a) and local delamination in panel (b) after 14 days *in vitro* culture. Gaps in the $SiO_2$ PECVD layer around the edges of the metal leads are also observed in panel (c), SEM image of a device with dual $SiO_2$/parylene C passivation. d–f) To increase adhesion with the top PECVD passivation layer, we added a 10 nm thin Ti layer on top of the Au metal leads. We also increased the thickness of the $SiO_2$ passivation layer to 500 nm. e-f) With these modifications, accelerated aging in DPBS solution at 60 °C demonstrated intact platform.

1.4 Surface roughness optimization

The USNW array devices were fabricated on the prime-grade Si substrate, and the surface of the devices (thermally grown $SiO_2$) showing a rms surface roughness less than 1 nm in the atomic force microscopy (AFM) topography image (Figure S6a). Since the surface roughness influences the cell's attachment and neurite formation, we optimized the surface roughness by two methods in Figure S6b,c. To create a rough surface on our device, we employed diluted HF (volume ratio of HF:DI water = 1:10) since it does not etch the surface $SiO_2$ uniformly. Our device had two types of $SiO_2$: the first layer is a thermally grown $SiO_2$ while second layer is PECVD grown $SiO_2$. First, we used diluted HF to etch the device with as-thermally grown 200 nm $SiO_2$ and the surface roughness was characterized by AFM (Figure S6b). The 30 s etching produced 2.22 nm average roughness and 20.26 nm maximum roughness; the 60 s etching created 1.83 nm average roughness and 15.20 nm maximum roughness; the 90 s etching created 0.37 nm average roughness and 4.29 nm maximum roughness. Then, we used diluted HF to etch the device with 500 nm $SiO_2$ deposited by PECVD and the surface roughness was characterized by AFM (Figure S6b). The 30 s etching produced 1.12 nm average roughness and 15.62 nm maximum roughness; the 60 s etching created 4.34 nm average roughness and 50.71 nm maximum roughness; the 90 s etching created 12.2 nm average roughness and 147.03 nm maximum roughness.

Earlier studies have shown that parylene C is a bio-compatible organic polymer material,[2] and parylene C with high nanoscale surface roughness and stable hydrophilic surface is good for protein attachment that helps cell adherence after plasma treatment.[3] We used parylene C as a top layer on our devices to preserve the cell's health during culture on our platform. 500 nm parylene C was deposited by SCS PDS 2010 Specialty Parylene Coating System. Then, we spin-coated a layer of positive photoresist (AZ 1518), applied a hollow polydimethylsiloxane (PDMS) mask with Al foil to protect the center parylene C layer and the USNW array during exposure step of the photolithography, followed by developing with AZ 300 MIF developer. The exposed parylene C at the pad regions and USNW tips were etched by $O_2$ plasma in Oxford P80 RIE etcher while the remaining part was protected by photoresist AZ 1518; after tip exposure, the passivating AZ1518 was removed. The final SEM image of the USNW with two passivation layers is shown in Figure S7a. We also used AFM to characterize the surface roughness, which showed an average roughness of 7 nm in Figure S7b.

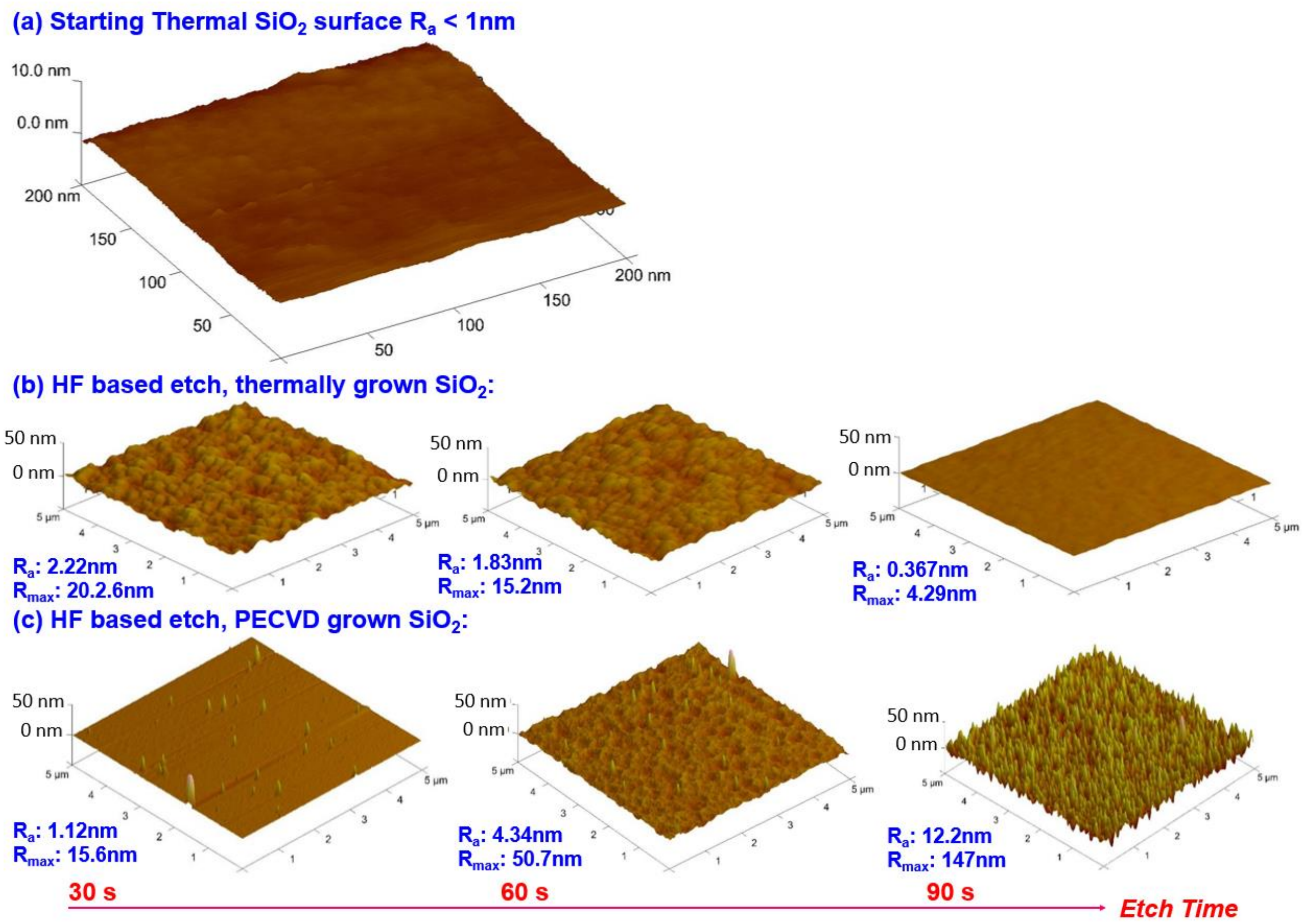

**Figure S6.** a) Atomic force microscopy (AFM) topography images of the surface of the devices (thermally grown $SiO_2$) showing a rms surface roughness less than 1 nm. PECVD deposited $SiO_2$ also exhibited a similar rms surface roughness. b) We used diluted HF treatment to roughen the surface to promote neuronal adhesion to the surface of the device. We observed that with the etching time increase, the rms surface roughness decreases. c) For PECVD grown $SiO_2$, the surface roughness increased with time. In our experiments, we used samples with rms surface roughness of 2 - 5 nm.

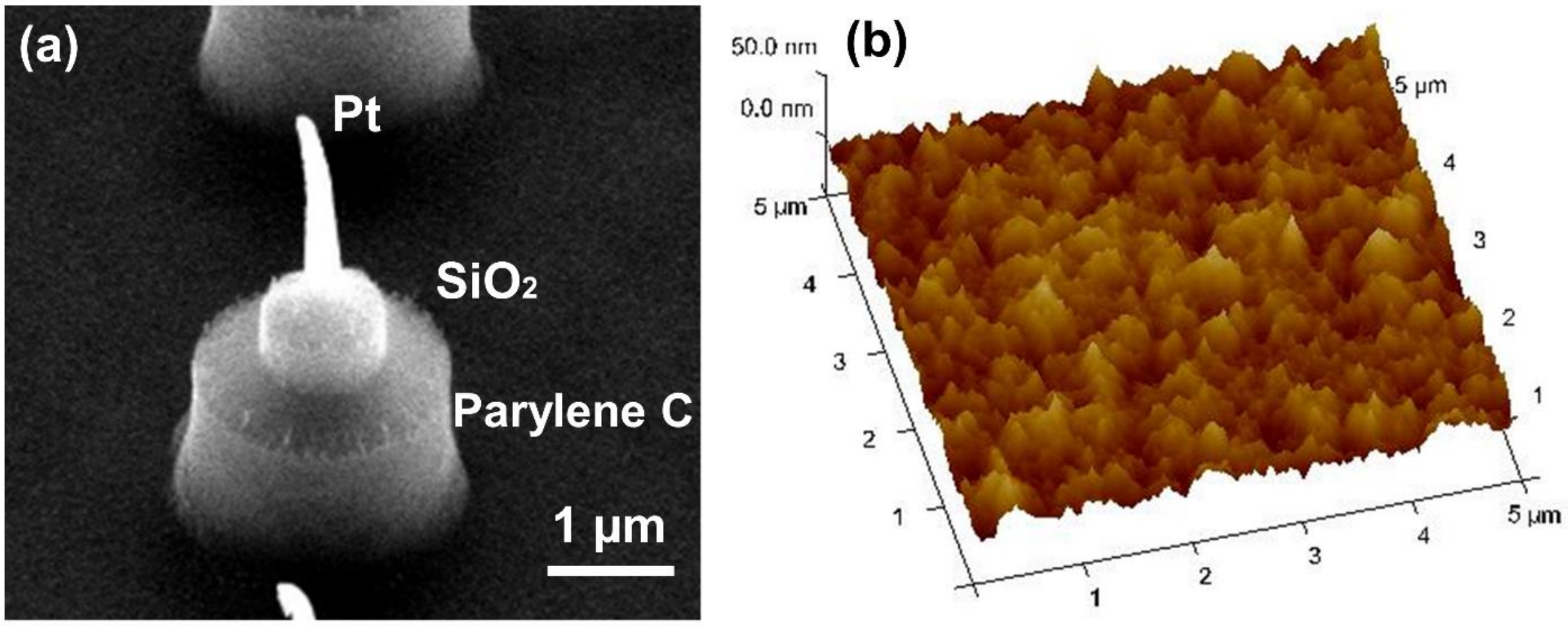

**Figure S7.** a) SEM image of a device with dual $SiO_2$/parylene C passivation. b) AFM image of dual $SiO_2$/parylene C passivated surface. We performed $O_2$ plasma treatment to roughen the surface of the parylene C, similar to the $SiO_2$ surfaces. The resulting rms surface roughness was 7 nm.

1.5 Packaging of the USNW array device and FFC bonding

Following fabrication, passivation and surface roughening, we bonded a culture chamber ring in the center of the devices using custom-made PDMS for sealing, where cells were cultured inside of the chamber. Then, the device was bonded with the flexible flat cable (FFC) by using the anisotropic conductive film (ACF) bonding at 100 °C (Figure S8).

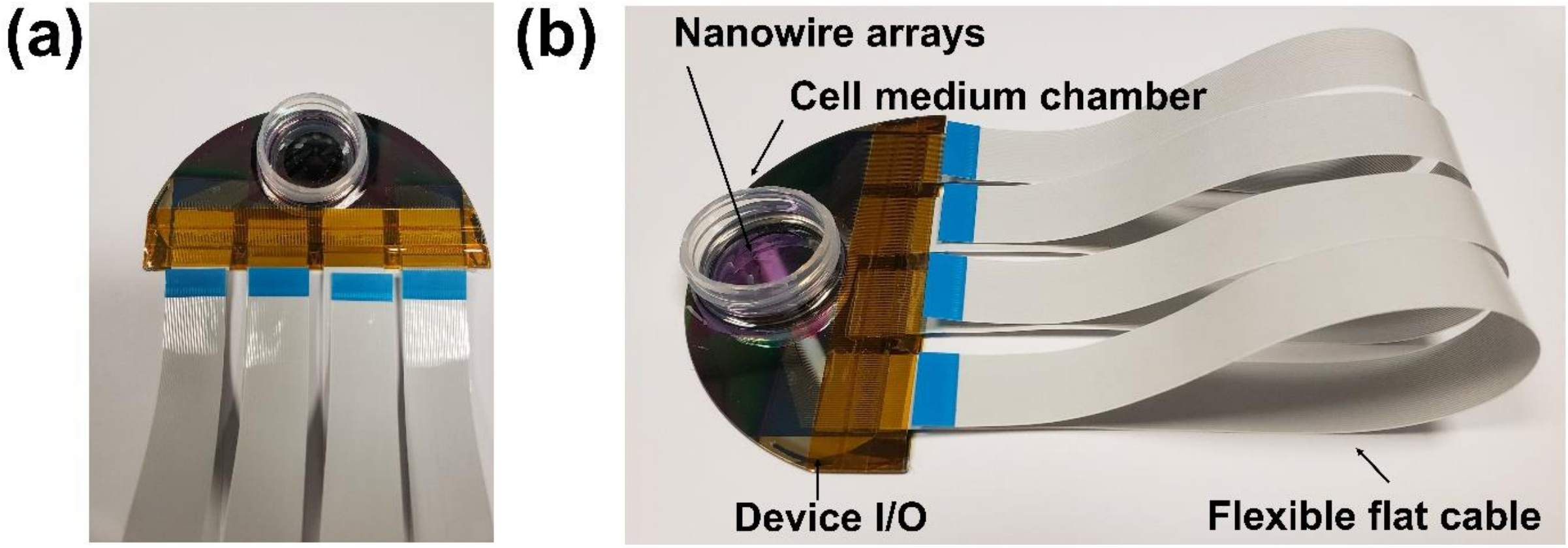

**Figure S8.** a-b) Picture of a competed device with a custom-made PDMS-sealed culture chamber and ACF bonding of FFC (6" long). The culture ring is adhered to the parylene C/ PECVD $SiO_2$/ thermal oxidation $SiO_2$/ Si-substrate surface with PDMS. The open central area for cell culture has an approximate diameter of 7 mm.

## 2. Electrochemical characterizations of individually electrically addressable Si USNW array devices

### 2.1 1 kHz electrochemical impedance assessment across arrays

The 1 kHz electrochemical impedance of the 128 channels of the USNW array devices was measured by the Intan RHS2000 Stimulation/Recording System by using a Pt wire as a counter electrode and soaking in the Dulbecco's phosphate-buffered saline (DPBS) solution. A representative of the 128 channel devices shows 100% yield based on impedance measurements (Figure S9a) and an average 1kHz impedance of 15.2±0.3 MΩ, with four arrays' average 1kHz impedance of 14.4±0.5 MΩ, 11.6±0.2 MΩ, 17.4±0.6 MΩ, 16.4±0.6 MΩ, respectively.

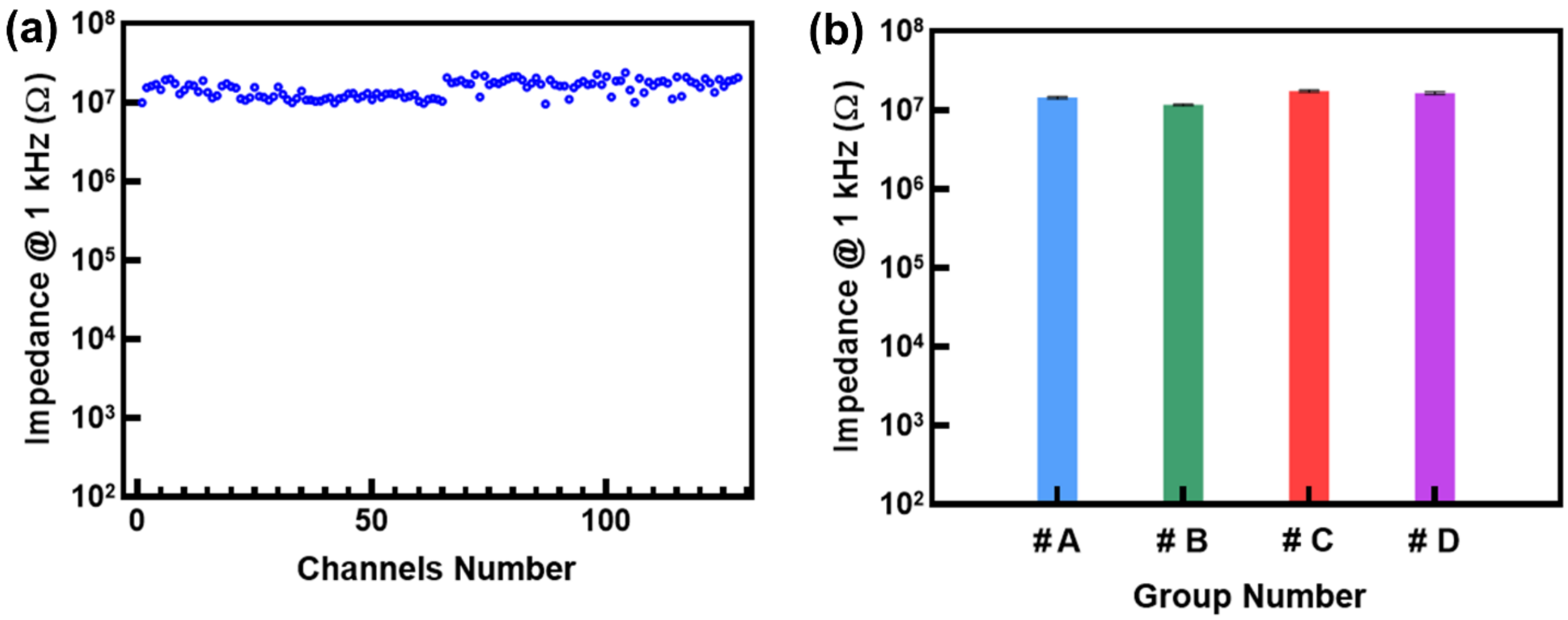

**Figure S9. a)** Electrochemical impedance of the 128 channels of individual USNW array at 1 kHz. **b)** Average electrochemical impedance at 1 kHz of each of 32 channels of individual USNW array in groups A, B, C, and D (14.4±0.5 MΩ, 11.6±0.2 MΩ, 17.4±0.6 MΩ, 16.4±0.6 MΩ, respectively).

2.2 Electrochemical impedance spectroscopy and charge injection capacity

To improve the electrical stimulation efficiency, we electroplated poly(3, 4-ethylenedioxythiophene):poly(styrenesulfonate) (PEDOT:PSS) on the stimulation channels of the array (Figure S10), since PEDOT:PSS allows higher current injection than the bare electrode. Next, electrochemical impedance spectroscopy (EIS) was performed by using the GAMRY Interface 1000E with merging the USNW array in DPBS solution and using the three electrode setup[2]: a large Pt electrode as a counter electrode, an Ag/AgCl electrode as a reference electrode, and USNW with electroplated PEDOT:PSS as the working electrode. Then, sinusoidal signals with 10 mV rms (root mean square) AC voltage (zero DC voltage) were applied. We measured the electrochemical impedance for a bare USNW (Figure S11a) and a USNW with electroplated PEDOT:PSS (Figure S11b) with the frequency range from 1 Hz to 10 kHz. The electrode impedance decreased by 50 times.

In addition, electrochemical current pulse injection with chronopotentiometry mode was performed by using the GAMRY Interface 1000E by submerging the USNW array in DPBS solution and using three electrodes setup: a large Pt electrode as a counter electrode, an Ag/AgCl electrode as a reference electrode, and USNW or USNW with electroplated PEDOT:PSS as the working electrode. Cathodic first, bi-phasic, charge-balanced current pulse was injected across the counter electrode and the working electrode, while the working electrode's polarization potential with reference to Ag/AgCl reference electrode was measured. $E_{mc}$ was calculated as working electrode potential versus Ag/AgCl reference electrode 10 μs after cathodal pulse phase, while $E_{ma}$ was calculated as working electrode potential versus Ag/AgCl electrode 10 μs after anodal pulse phase. For organic electrodes (PEDOT:PSS/Au or PEDOT:PSS/Pt), the water window limit is between -0.9 V to 0.6 V; while for metallic electrodes (Pt or Au), water window limit is between -0.6 V to 0.8 V.[4] $E_{mc}$ is the potential when the electrode/solution interfacial potential reaches reduction (cathodal limit) and $E_{ma}$ is the potential when it reaches oxidation (anodal limit). Charge injection capacity (CIC) is the total charge density at which either $E_{ma}$ reaches water oxidation potential or $E_{mc}$ reaches water reduction potential expressed by $CIC=Q_{inj}/GSA$ (geometric surface area). The CIC for bare Pt USNW and the USNW with electroplated PEDOT:PSS were 0.60 mC $cm^{-2}$ and 2.98 mC $cm^{-2}$, respectively (Figure S11c,d). Electroplating PEDOT:PSS on the USNW increases its current injection capacity by nearly 5 times. The PEDOT:PSS is therefore more effective for neuronal stimulation to avoid reaching water hydrolysis.

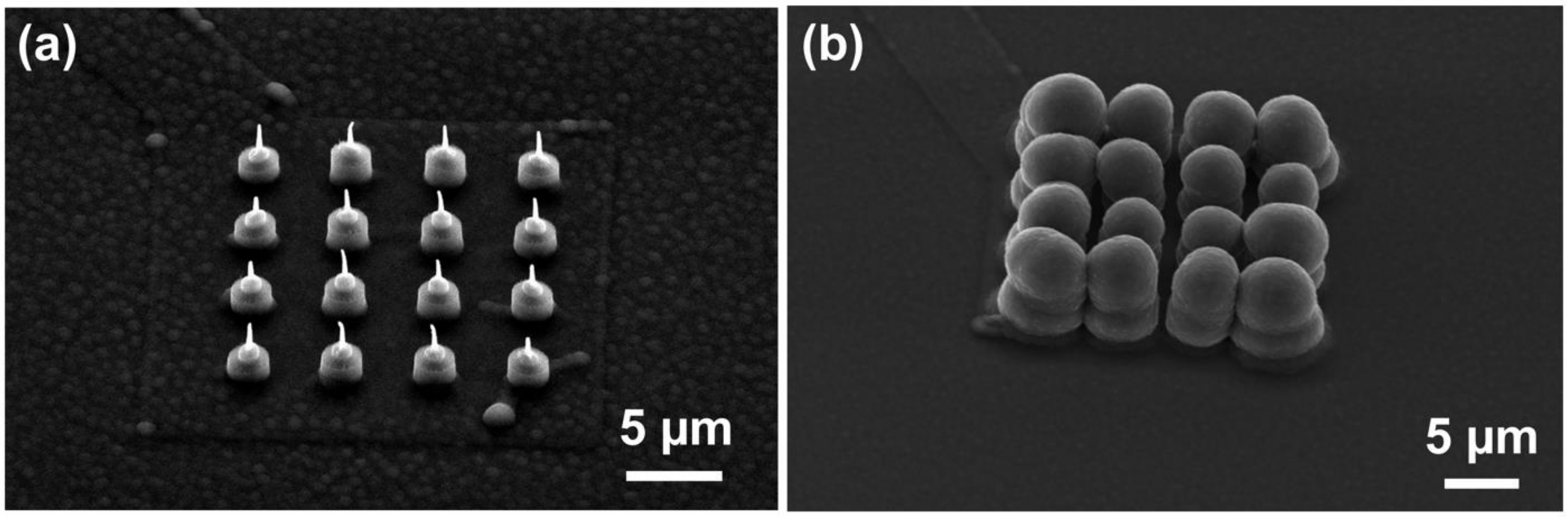

**Figure S10.** a-b) SEM image of a pad with 16 NWs (a) before and (b) after PEDOT:PSS electroplating.

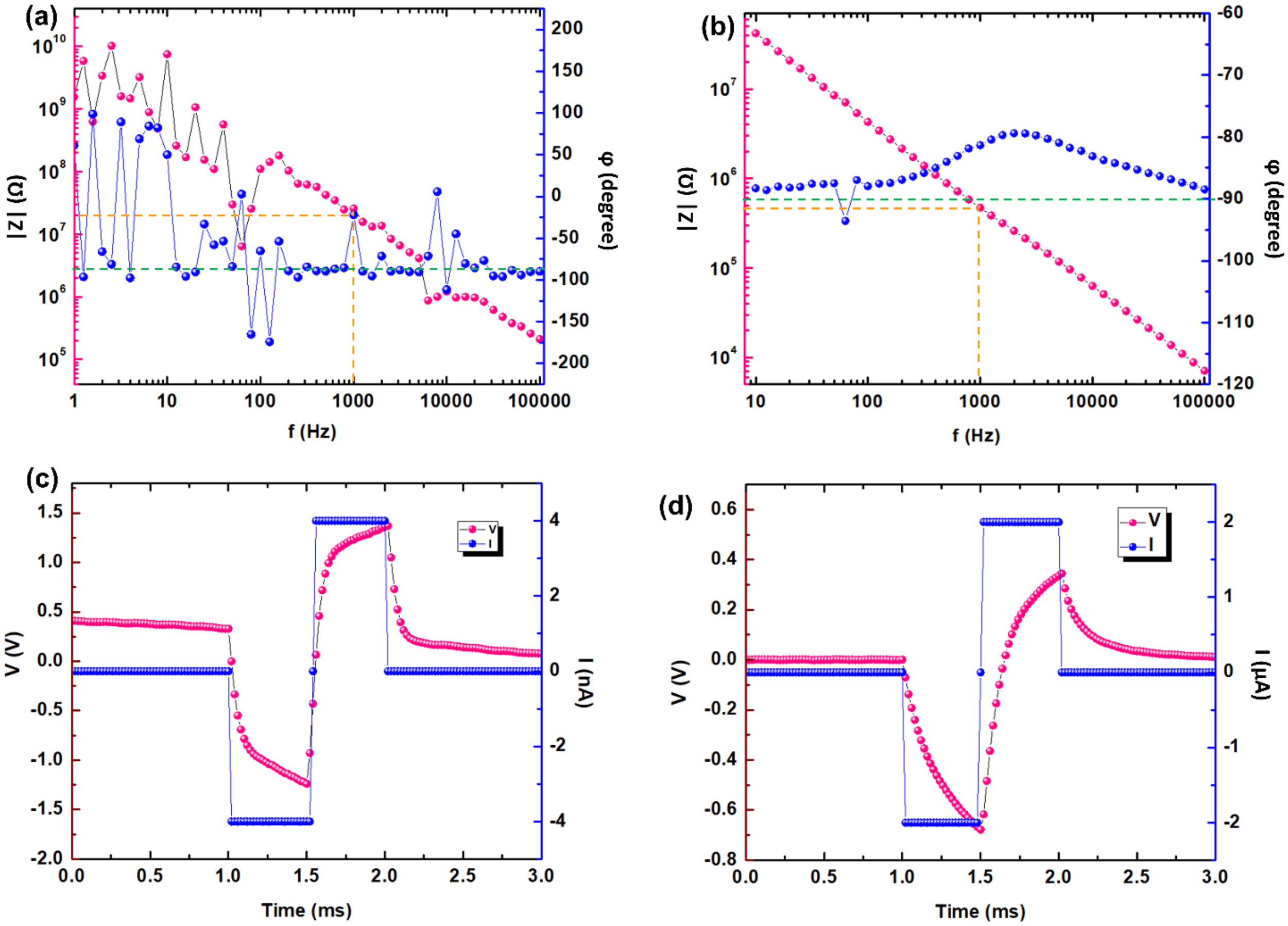

**Figure S11.** a-b) Impedance spectroscopy for (a) 16 NWs with bare Pt NW surface and (b) after PEODT:PSS electroplating, showing lowered impedance by ~50 times. c-d) Voltage transients at the water hydrolysis window for bare NW with Pt surface (c) showing a maximum current capability of 0.60 mC $cm^{-2}$ and (d) after PEDOT:PSS electroplating showing a maximum current capability of and 2.98 mC $cm^{-2}$ before breaking water.

**3. Cell culture**

3.1 Rat cortical neuron culture on USNW array devices

The integration of USNW array electrodes with cells followed immediately from fabrications and packages of the device, which involved the following steps (steps 1 to 4 were also applied before iPSC-CVPCs cultures):

(1) The USNW array device was sterilized by 1) DI water rinse for three times, and 2) 70% ethanol for at least 30 min.

(2) The device was washed with PBS for three times followed by placing 20 μl (0.1% w/v) Polyethylenimine (PEI, Sigma-Aldrich) solution on the USNW array and incubating them at 37 °C incubator for 1 hour.

(3) The PEI was aspirated and washed with 500 μl double distilled water (dd$H_2O$) for 4 times. The device dried in the incubator for 5 hours or overnight.

(4) 20 μl laminin (20 μg $ml^{-1}$, Sigma-Aldrich) was added to the USNW array spot and incubated at 37 °C incubator for 1 hour.

(5) Medium (Neurobasal (Thermo Fisher Scientific) +2% B27 (Thermo Fisher Scientific) +1% P/S (Corning) + 10% FBS) was prepared for plating rat cortical neurons (RCNs) on USNW arrays.

(6) The cryopreserved rat neurons cryovial (from Thermo Fisher Scientific) was removed from the liquid nitrogen storage container, warmed in a 37 °C water bath for exactly 2.5 min, sprayed the outside with 70% ethanol, wiped dry, and placed in a bio-safety cabinet.

(7) The contents of the cryovial were carefully transferred to a 15 mL centrifuge tube using a 1 mL pipettor. The inside of the cryovial was carefully washed with 1 mL of room temperature Neurobasal medium (~1 drop $s^{-1}$). 3 mL of room temperature Neurobasal medium was slowly added to the tube (~1-2 drops $s^{-1}$). The contents were carefully mixed by inverting the tube 2-3 times. The total number of cells in suspension was determined via hemocytometer count.

(8) The cells were concentrated by centrifuging 1100 rpm for 3 min. Laminin was aspirated and 20 μl cell suspension was plated directly on the USNW arrays with the cell density at 200k cells/array. The USNW devices with seeded neurons were incubated in a cell culture incubator at 37 ℃, 5% $CO_2$ for 35-40 min. Next, 500 μL of warm Neurobasal medium was carefully added to chamber from the side of the device. To avoid adding the medium too fast to cause the detachment of the adhered neurons on the USNWs, the USNW devices with seeded neurons were put back to the cell culture incubator at 37 ℃, 5% $CO_2$ for another 35-40 min. Then, another 500 μL of warm Neurobasal medium was carefully added to chamber from the side of the device to reach a volume of 1.5 mL per device.

(9) On the next day, the medium was changed to Brainphys Complete (Stem cell) + supplements (brain-derived neurotrophic factor (BDNF), glial cell line-derived neurotrophic factor (GDNF), Ascorbic acid, cyclic adenosine monophosphate (cAMP) and Laminin). Then, half the medium was replaced with fresh medium for every other day with Brainphys + supplements till 7 days *in vitro* (DIV), then switched back to Brainphys Complete with no added supplements. This showed clear neuronal network formation on the USNW array platform (Figure S12).

In parallel to plating on USNW array devices, rat cortical neurons were also plated at a density of 20,000 cells/well in (Neurobasal (Thermo Fisher Scientific) +2% B27 (Thermo Fisher Scientific) +1% P/S (Corning) + 10% FBS) on 384 well imaging plates (Poly-D-lysine treated, Biocoat, Corning), which had been pre-coated with laminin (20 ul/well) for 1h at 37oC. Cells were maintained in a humidified 37°C incubator with 5% CO2, with media exchanges into BrainPhys that exactly paralleled those described above for culture on USNW array devices.

Rat cortical neuronal cultures on 384-well imaging plates were fixed in 4% paraformaldehyde (PFA) (Alfa Aesar Chemicals) for 10 min at room temperature at 2 and 4 weeks after plating. The cells were then washed 3 times in phosphate buffered saline (PBS) (Gibco|Thermo-Fisher Scientific), and then blocked in 5% donkey serum (Jackson ImmunoResearch) and permeabilized in 0.1% Triton-X (Sigma-Aldrich) in PBS. Primary antibodies at the dilutions noted (Table 1) were incubated on the cells overnight at 4ºC. The following day cells were washed 3 times in PBS, and the secondary antibodies (AlexaFluor, Molecular Probes) were added 1:500 in blocking buffer for 2 hours at room temperature. The secondary antibody was then washed several times in PBS and DAPI (Thermofisher, 1:2000 in PBS) was added for 30 min at room temperature. Images were acquired with the Opera Phenix High Content Screening System confocal microscope (Perkin Elmer).

| **Antibody** | **Species** | **Dilution** | **Company** |
|---|---|---|---|
| GFAP | Goat | 1:500 | Santa Cruz Biotechnologies |
| MAP2 | Guinea Pig | 1:1000 | Synaptic Systems |
| Tuj1 | Mouse | 1:500 | Biolegend |

**Table S1.** Antibodies Specification.

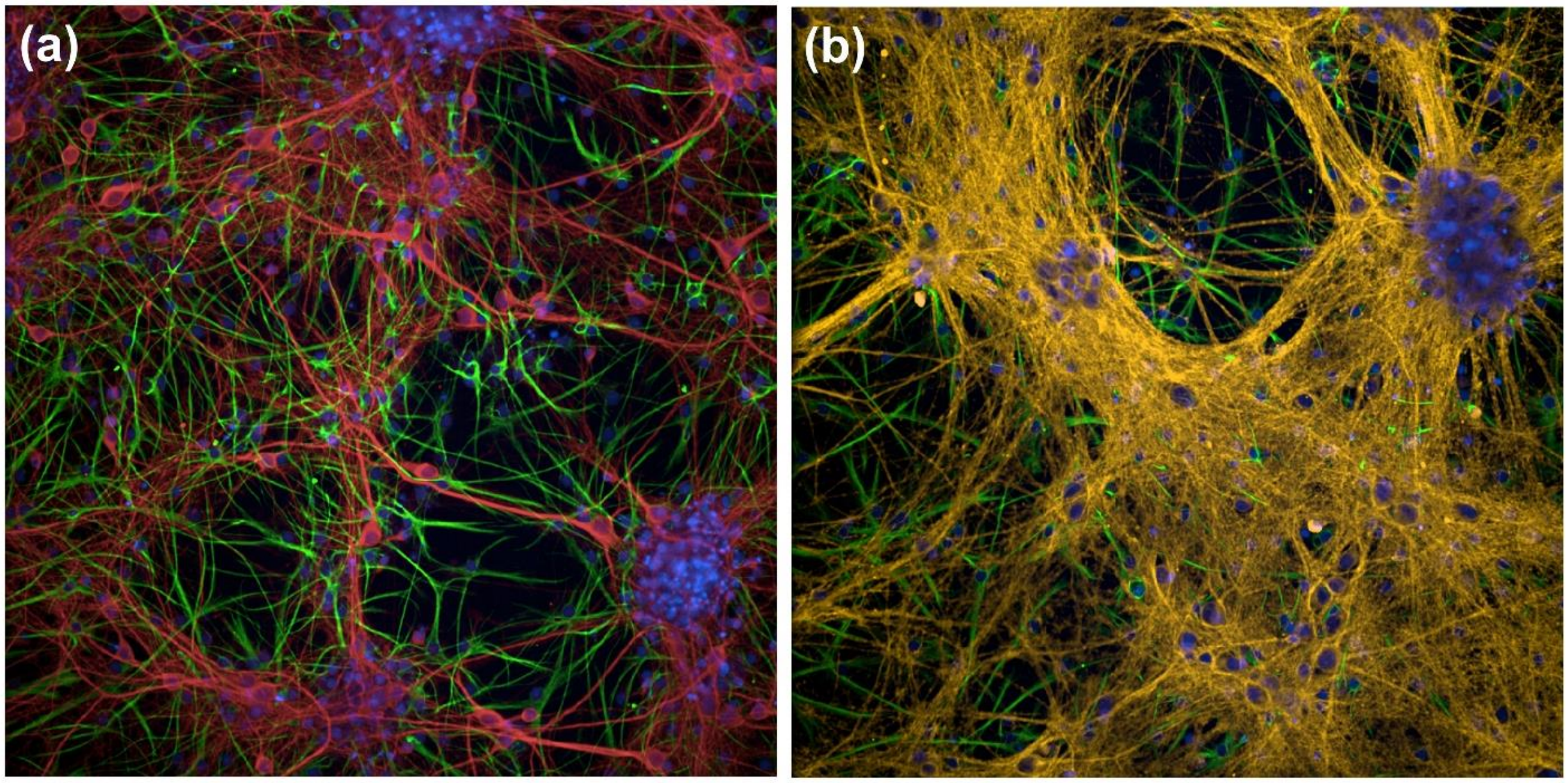

**Figure S12.** a-b) Fluorescence images of rat cortical neurons (a) Red: MAP2b neuronal marker; Green: GFAP, astrocyte marker; Blue: DAPI nuclear marker. (b) Yellow: β-Tubulin, neuronal marker; Green: GFAP, astrocyte marker; Blue: DAPI nuclear marker.

3.2 iPSC-CVPCs culture on USNW array devices iPSC-derived cardiovascular progenitor cells (iPSC-CVPCs) were derived and cryopreserved at day 25 (D25) of differentiation.[5, 6] The cryopreserved iPSC-CVPCs were thawed either on a 6-well plate coated overnight with Matrigel (Corning) or directly on a USNW device coated with Matrigel for 5 hours or overnight. The following day, the medium was changed with fresh iPSC-CVPC medium (RPMI 1640 (Thermo Fisher Scientific) containing B27 Supplement (50X) (Thermo Fisher Scientific) and Pen-Step (Thermo Fisher Scientific)). When the iPSC-CVPC were thawed onto a 6-well plate, the cells were cultured for 2 days before plating on a USNW device. $5 \times 10^4$, $1 \times 10^5$ or $2 \times 10^5$ cells were plated over the active regions of a USNW device in 5 - 8 µL of the iPSC-CVPC medium containing B27 Supplement and Pen-Step (Thermo Fisher Scientific). After cells were plated on a USNW device, iPSC-CVPCs were cultured for 5 - 7 days prior to first recording with medium changed every other day. The day before recordings of iPSC-CVPCs, the medium was replaced with fresh iPSC-CVPC medium.

3.3 Pharmacological stimulation and inhibition

Pharmacological responses were assessed to validate the neuronal activities are of electrophysiological origin. Picrotoxin (PTX) is an equimolar mixture of two compounds, picrotoxinin (C15H16O6; CAS# 17617-45-7) and picrotin (C15H18O7; CAS# 21416-53-5). If neurons are under $GABA_A$ receptor-mediated inhibition, PTX will stimulate neurons by relief of inhibition.[7] Tetrodotoxin (TTX) is a sodium gate channel blocker which suppresses cell firing. We recorded the baseline activities of rat cortical neurons on 7 DIV in Figure 2m, which showed some action potentials on channel A003. Then, we added PTX to a final concentration of 33 nM to the culture medium, and the neuronal activity increased and generated high-frequency action potentials (Figure 2n). Next, TTX was added to a final concentration of 1 nM to the culture medium and we observed that the activity decreased until it disappeared (Figure 2o).

**4. SEM and FIB-SEM imaging on neurophysiology platform**

The USNW array devices with cells were fixed, dehydrated, critical point dried and coated with IrOx for SEM imaging. The fixative mediums were the mixes of distilled water, sodium cacodylate buffer (0.2 M, pH 7.4) and Glutaraldehyde solution (G58882-10X1ML, Grade I, 25% in $H_2O$) with the ratio of 4:5:1. We first melted Glutaraldehyde solution at 37 °C water bath, then these mediums were mixed together by using spinner of VORTEX-GENIE 2 for 3 mins. Firstly, the growth media was rinsed with PBS for three times. Subsequently, the samples had undergone a cell fixation protocol, in which a solution containing 2.5% glutaraldehyde in 0.1 M cacodylate buffer at pH = 7.4 was added. The sample sat for 1 hour at room temperature and was then washed three times in PBS with leaving the sample in PBS solution for 5 min after each rinse. Buffer salts were rinsed off with three washes in distilled water, 5 min for each rinse. The samples were then subjected to a dehydration procedure in which they were serially dehydrated in 30, 50, 70, and 90% (10 min each) and three times with 100% ethanol. Following the dehydration procedure, the samples were dried in critical point dry (CPD) by using Autosamdri-815 for ~10-15 min. Finally, a 10 nm IrOx was sputtered on the surface of the sample by using Emitech K575X coater (5-7 s at 85 mA) for SEM imaging (Neurons: Figure S13 and Figure S14, iPSC-CVPCs s: Figure S17) and 100+ nm of Ti was deposited on samples for protection of the surface for FIB processing (Figure S15 and Figure S16).

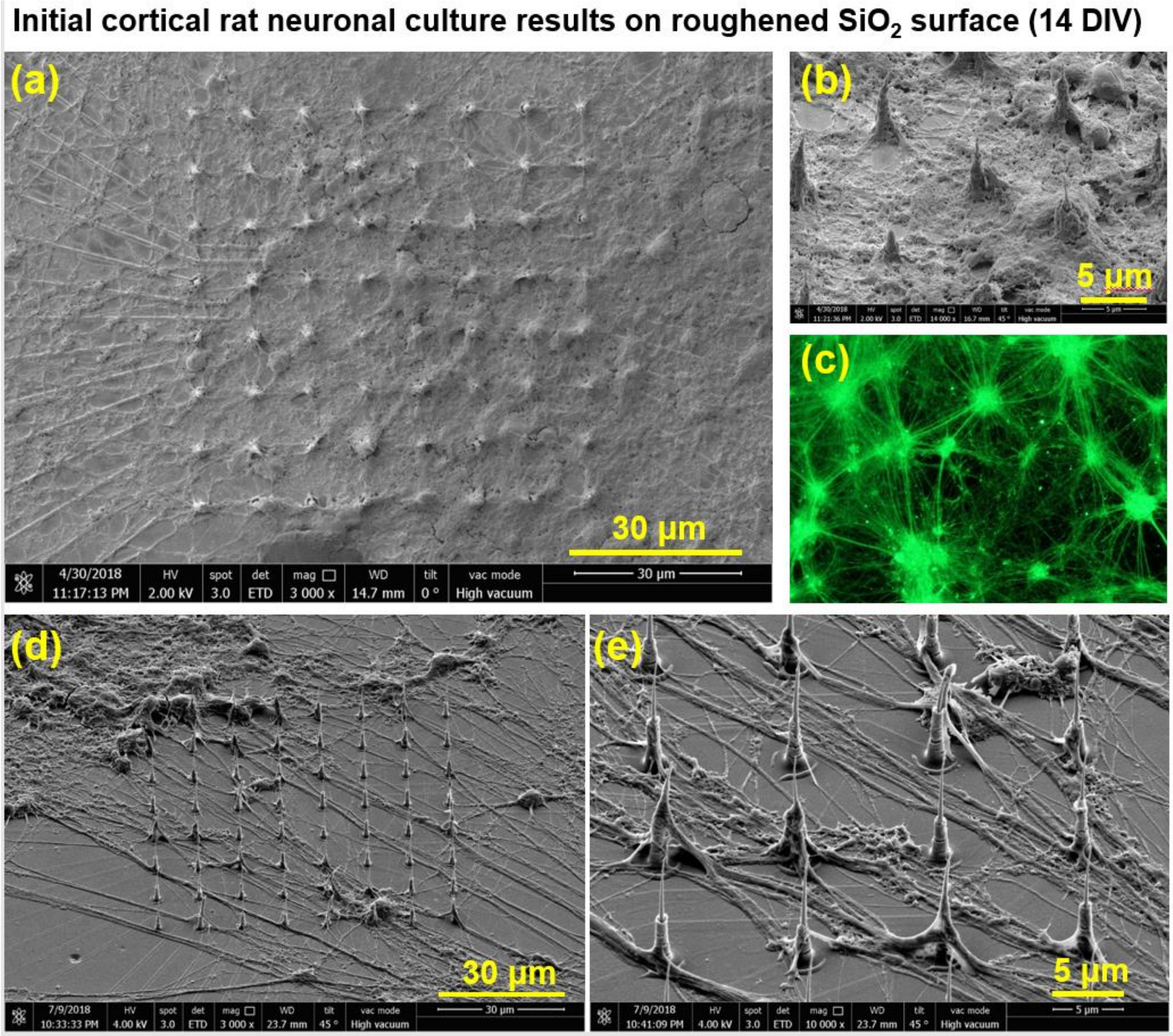

**Figure S13.** a, b, d, e) Top and angled view SEM images of cultured rat cortical neurons 14 days DIV on $SiO_2$-passivated USNW platform showing neurite extensions indicating healthy cell culture. **c,** Fluorescence imaging of the simultaneously cultured neurons on a control glass plate also demonstrating appropriate culture conditions. Total number of neurons was 200k for this experiment.

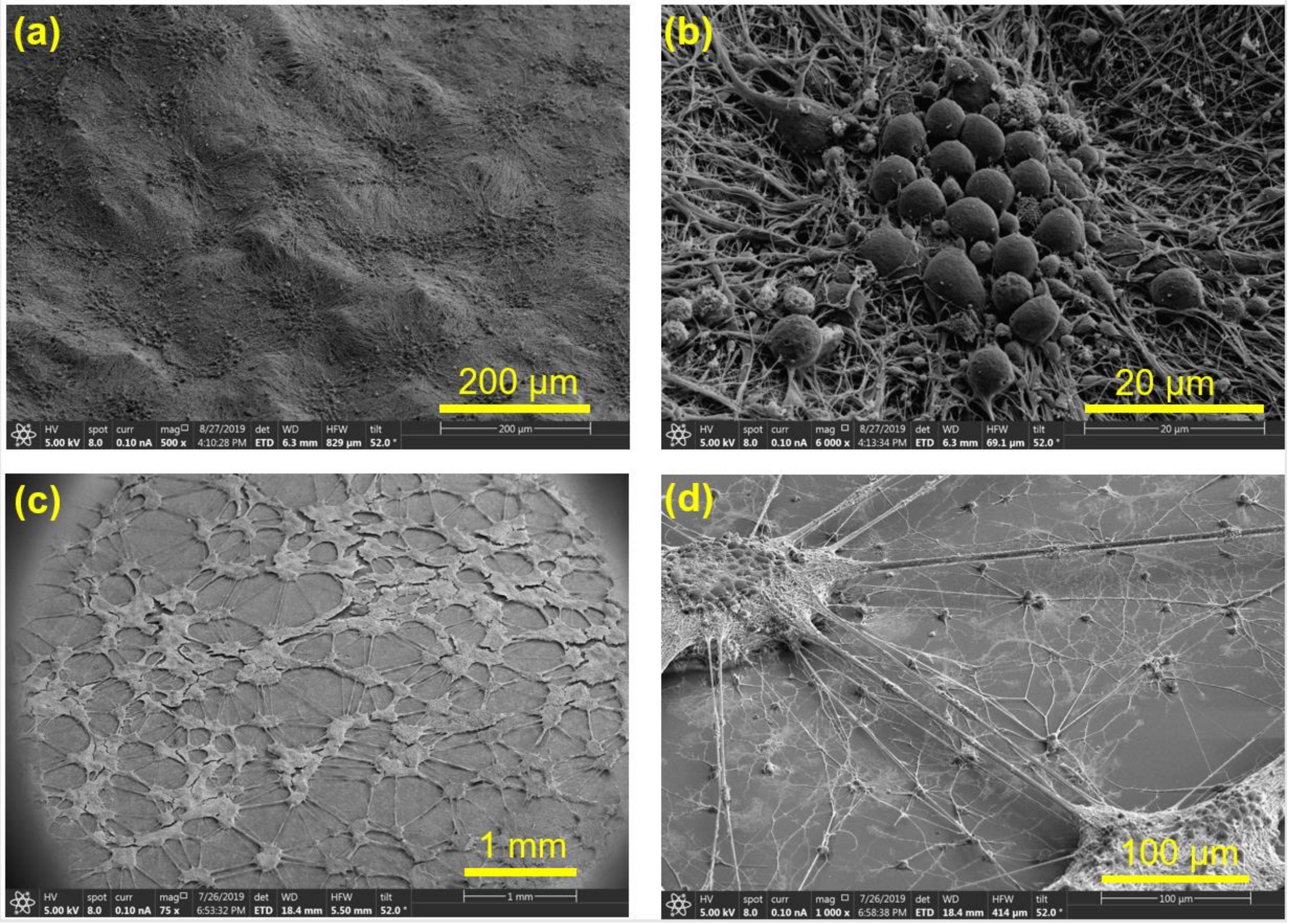

**Figure S14.** a-d) SEM images obtained from the cell culture on a single device. On certain portions of the device, we obtained continuous, tissue-like cell culture (a-b) as is further illustrated in Figure S8. In some other regions where no global clustering was observed, we obtained satellite-like clusters of 3D cultured layers (c-d). Total number of neurons was 800k for this experiment.

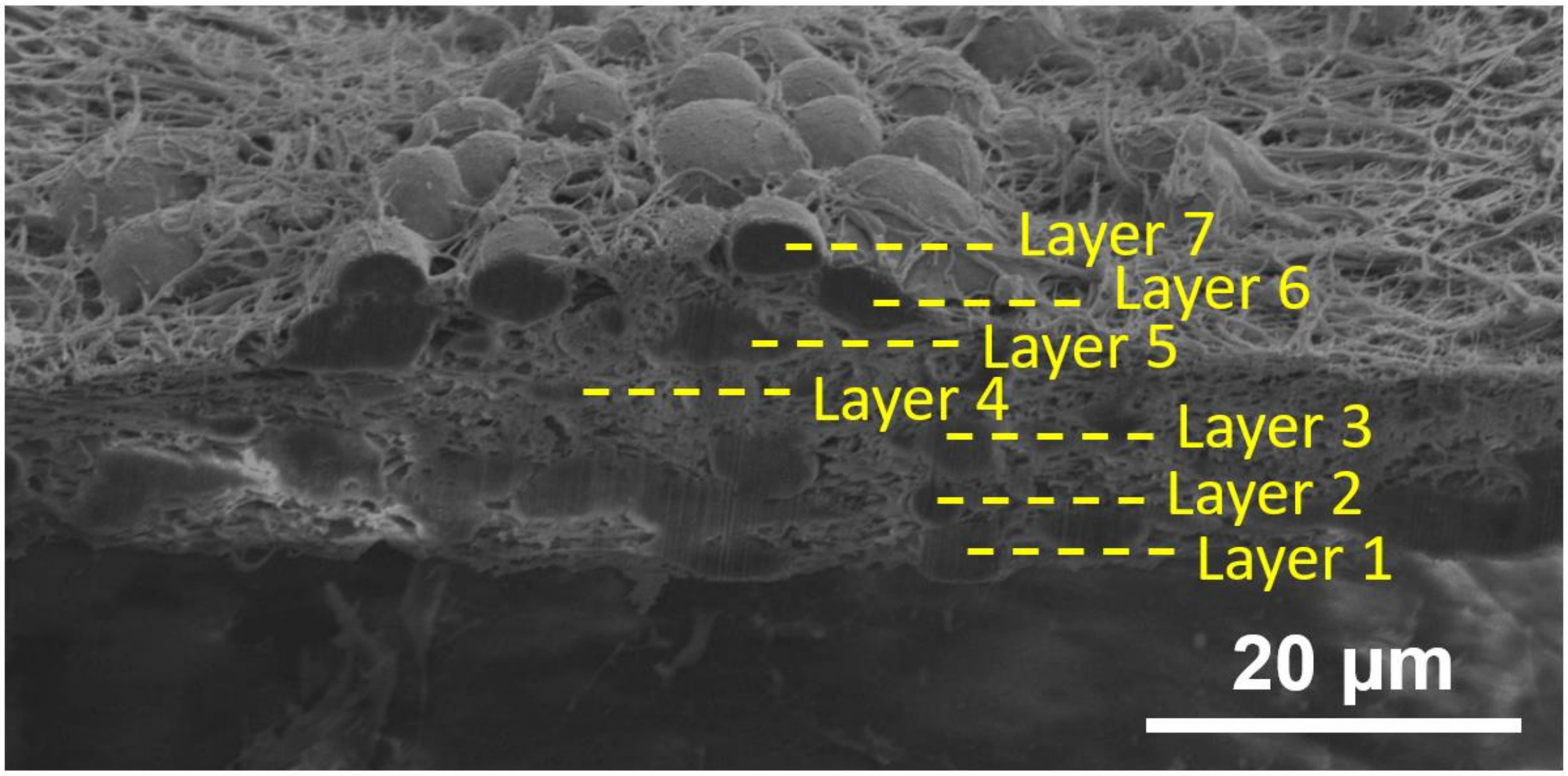

**Figure S15.** Original SEM image taken in the cross-section (sectioned by FIB) showing multi-layered neuronal structure.

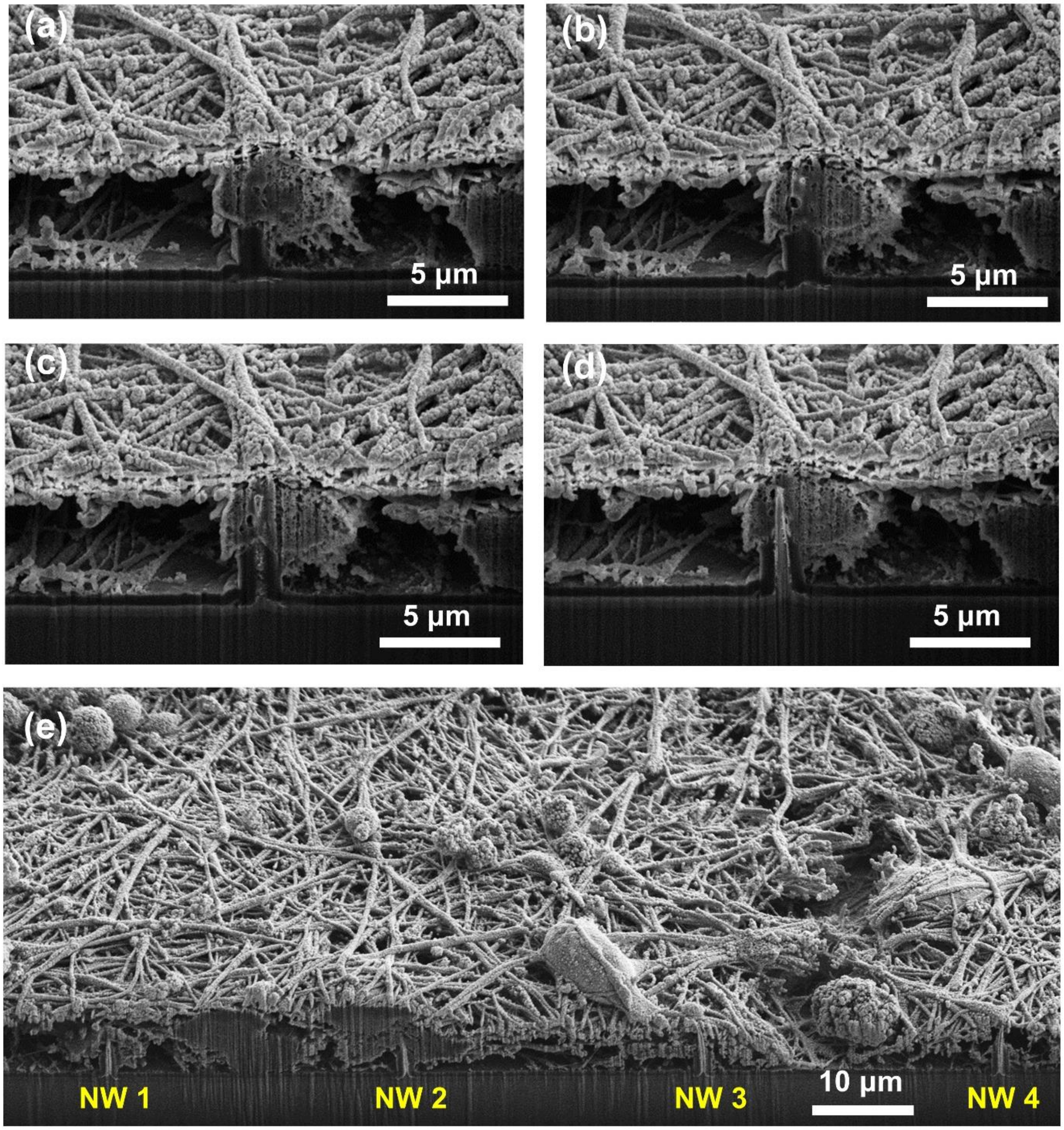

**Figure S16.** Original SEM image taken in the cross-section (sectioned by FIB) showing the interface between neurons and USNWs. a) Sequential FIB sectioning showing the interface of neuronal soma and edge of USNW, b) Sequential FIB sectioning revealing first the tip of the USNW inside the soma, c) Sequential FIB sectioning revealing part of the USNW inside the soma and d) the whole USNW/neuron cross-section, full SEM image showing clear penetration of USNW electrode into the neuron soma, and e) cross-section showing multiple USNW penetrations into the cultured neurons.

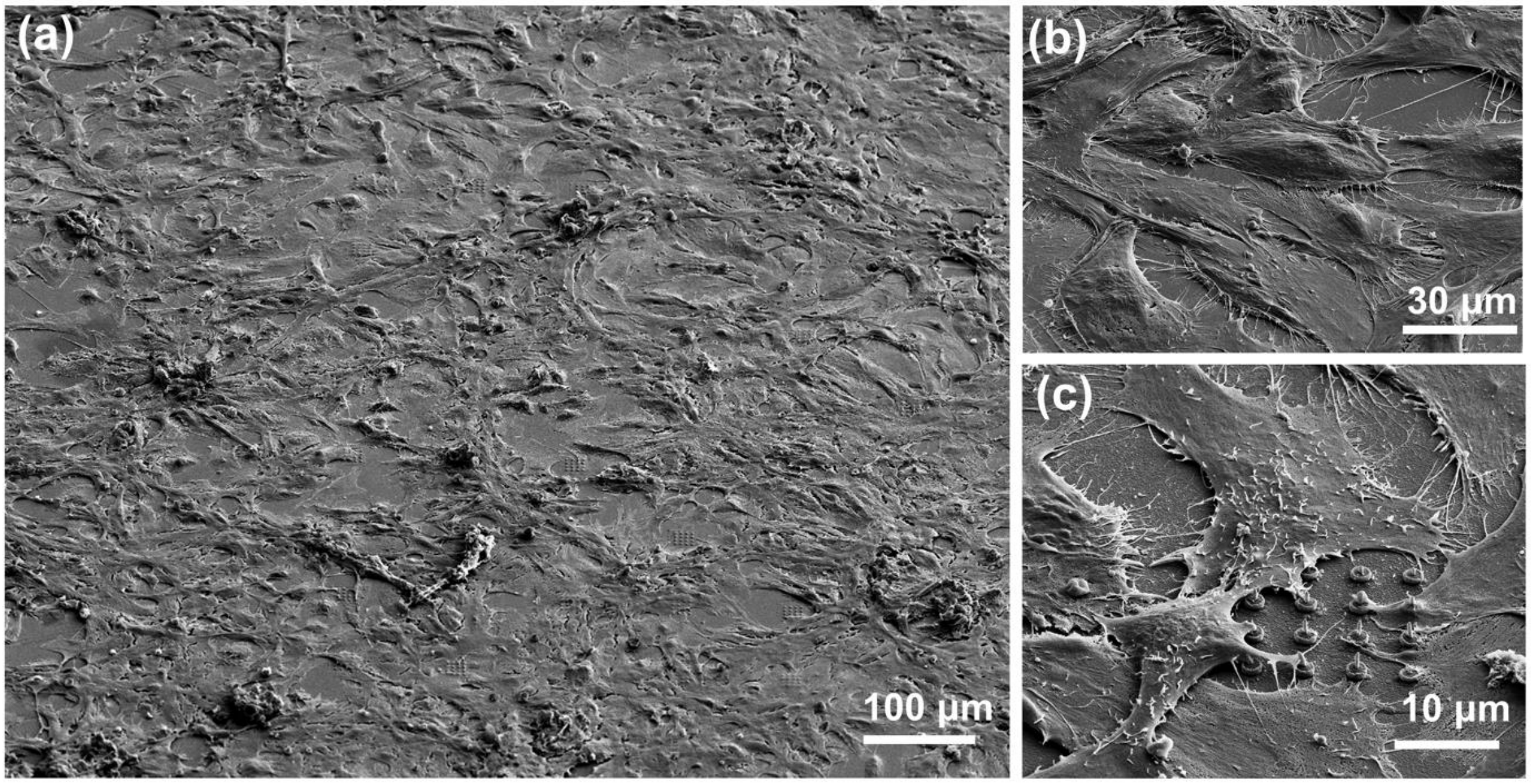

**Figure S17.** SEM images of cultured iPSC-CVPCs on the USNW platform: a) overview SEM image of the iPSC-CVPCs on the USNW arrays. b) Zoomed-in SEM image of the iPSC-CVPCs. c) Zoomed-in SEM image of iPSC-CVPCs on the USNW arrays.

## 5. Electrophysiological recordings of neuronal and cardiomyocyte activity

Multiple arrays with variable USNW pitches (5 µm, 10 µm, 30 µm, and 70 µm) were investigated where each array was composed of 32 USNWs. Electrophysiological recordings were carried out starting at 7 DIV for neurons (Figure S18) and at 5-7 DIV for iPSC-CVPCs s (Figure S23 and Figure S24) using Intan RHS2000 stim/recording system. A Pt wire was used as the ground and reference electrodes, and a sampling rate of 30000 samples $s^{-1}$ was used for the recording. The biphasic-pulse stimulation peak width, amplitude and frequency were 0.5 ms, 10 nA and 1 Hz, respectively. Matlab codes provided by Intan Technologies were used to convert raw data files into accessible format and post-processing.

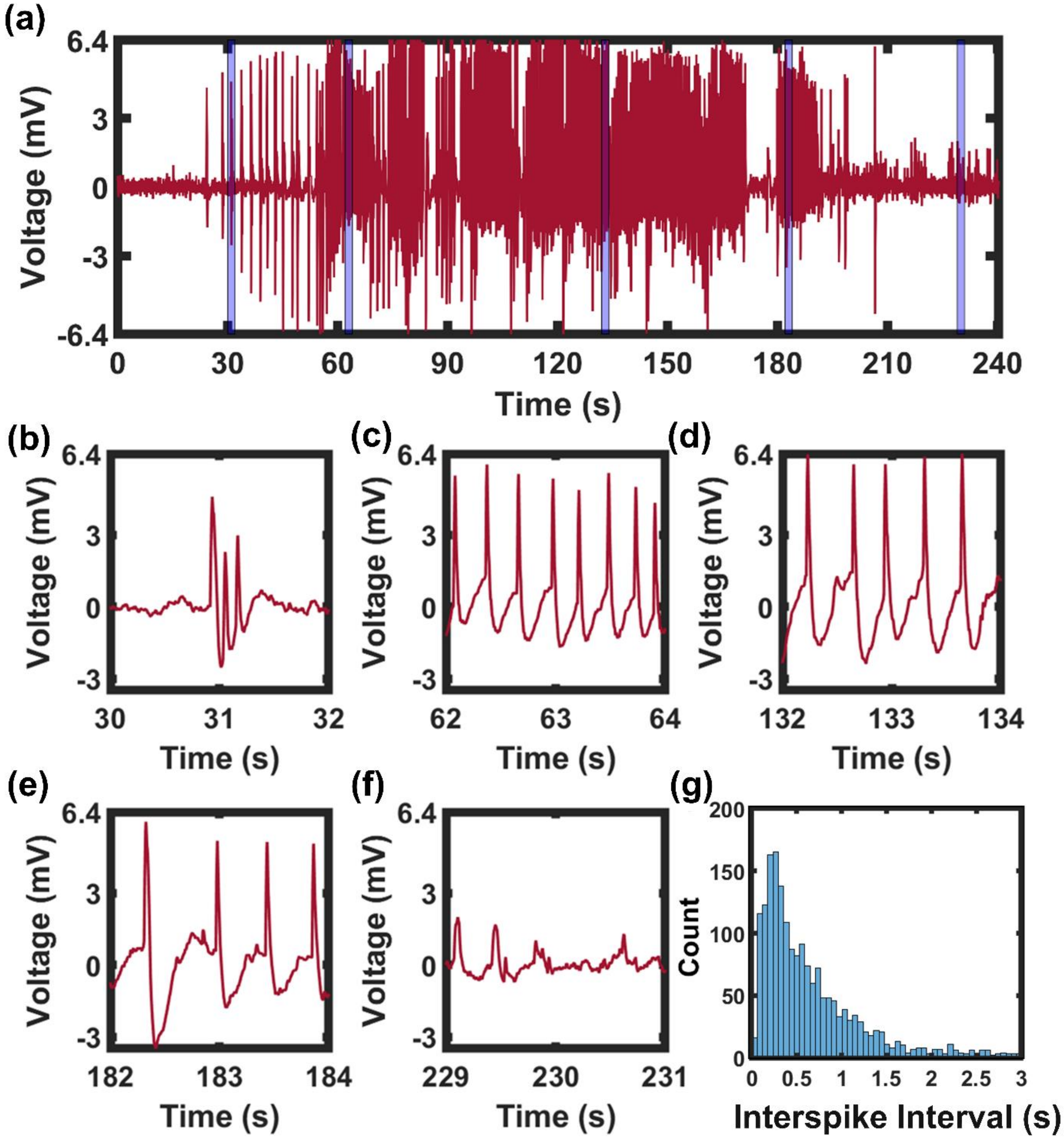

**Figure S18.** Single USNW recording segment demonstrating consistent signal amplitude over time and interspike analysis a) Continuous, 240 s recording segment from channel 31 USNW from group B (10 µm electrode spacing) at 11 DIV, showing consistent amplitude without attenuation. b-f) Close-up image of selected time segment over 240 s. The large amplitude action potential shapes c-e) throughout the recording duration remain consistent. g) The interspike interval width has an exponentially decaying tail, corresponding to action potential refractory periods and spike triggering from random processes.

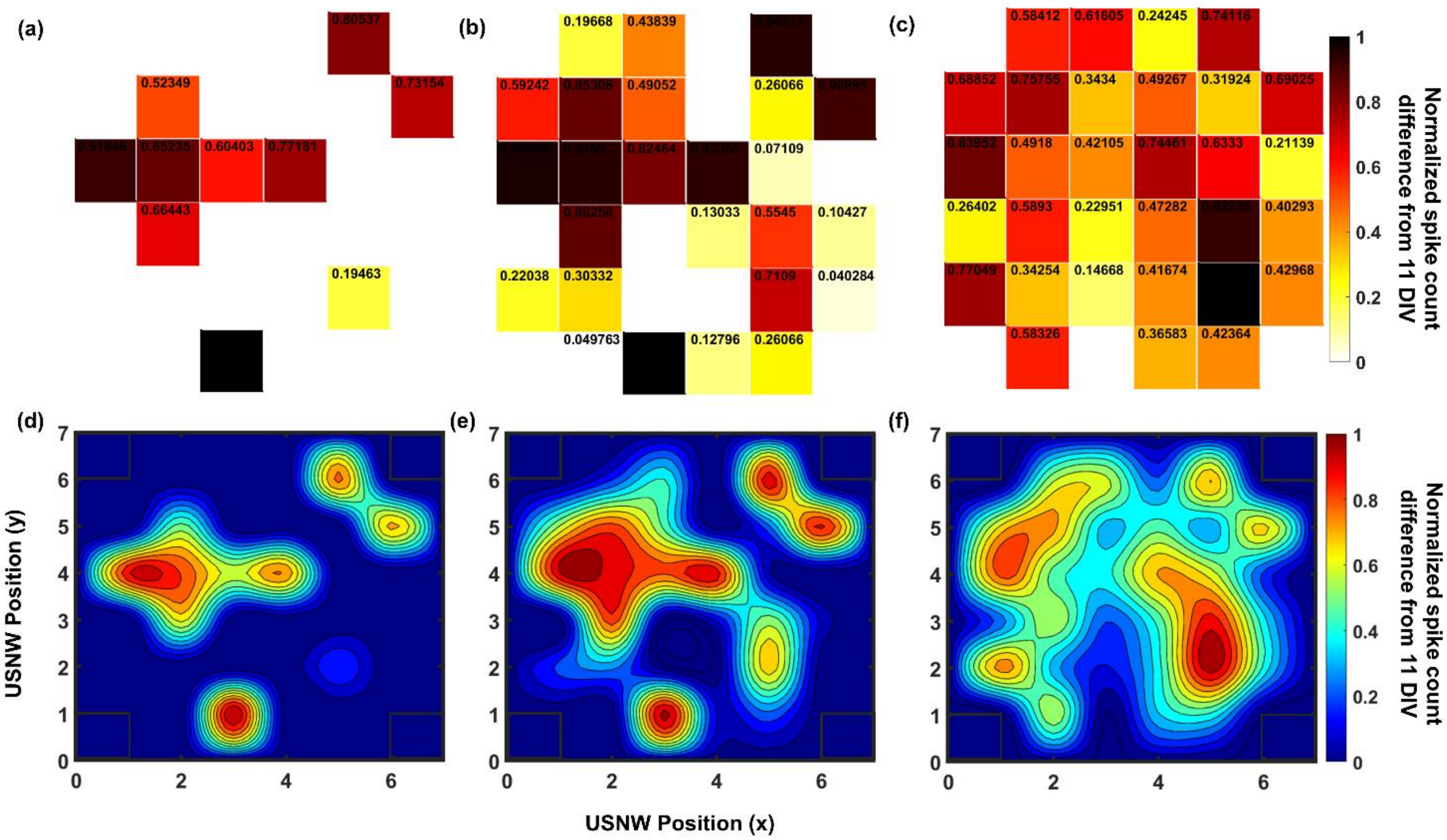

**Figure S19.** Normalized spike activity differences of USNW channels between 13, 15, and 19 DIV with 11 DIV. a) Normalized spike count difference from 11 DIV and 13 DIV (normalized to the maximum spike activity difference observed in the particular recording day across 32 channels). White square spaces indicate no increase in spike activities from 11 DIV. b) Normalized spike count difference from 11 DIV and 15 DIV. c) Normalized spike count difference from 11 DIV and 19 DIV. d-f) Contour map of normalized spike activity differences of 32 channels from 13, 15, and 19 DIV with spike activities from corresponding channels at 11 DIV, group B. The contour maps were plotted with 10 layers (10 channels with increased activity at 13 DIV from 11 DIV) and the 2D mesh grids were interpolated to provide more smooth contours. As the cells are cultured longer, we gradually see increase in spike activities across all 32 channels.

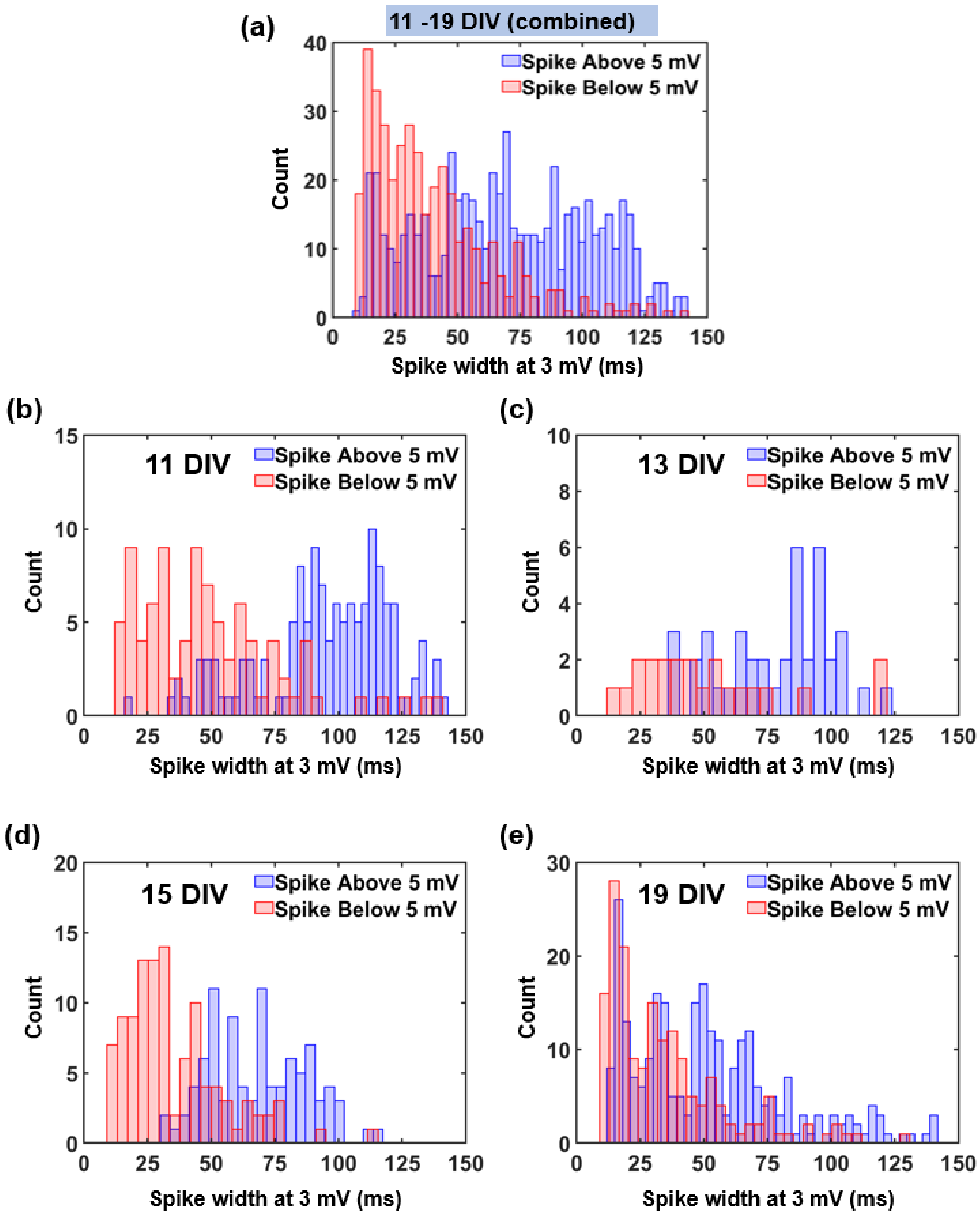

**Figure S20.** Histogram plot of spike width at 3 mV to compare spikes above and below 5 mV by days of culture. a) Combined histogram plot of spike width from 11 to 19 DIV. b) Histogram plot of spike width at 11 DIV (~360 detected spikes above 5 mV, ~90 spikes below 5 mV). c)

Histogram plot of spike width at 13 DIV (~40 detected spikes above 5 mV, ~20 spikes below 5 mV). d) Histogram plot of spike width at 15 DIV (~90 detected spikes above 5 mV, ~100 spikes below 5 mV). e) Histogram plot of spike width at 19 DIV (~520 detected spikes above 5 mV, ~170 spikes below 5 mV).

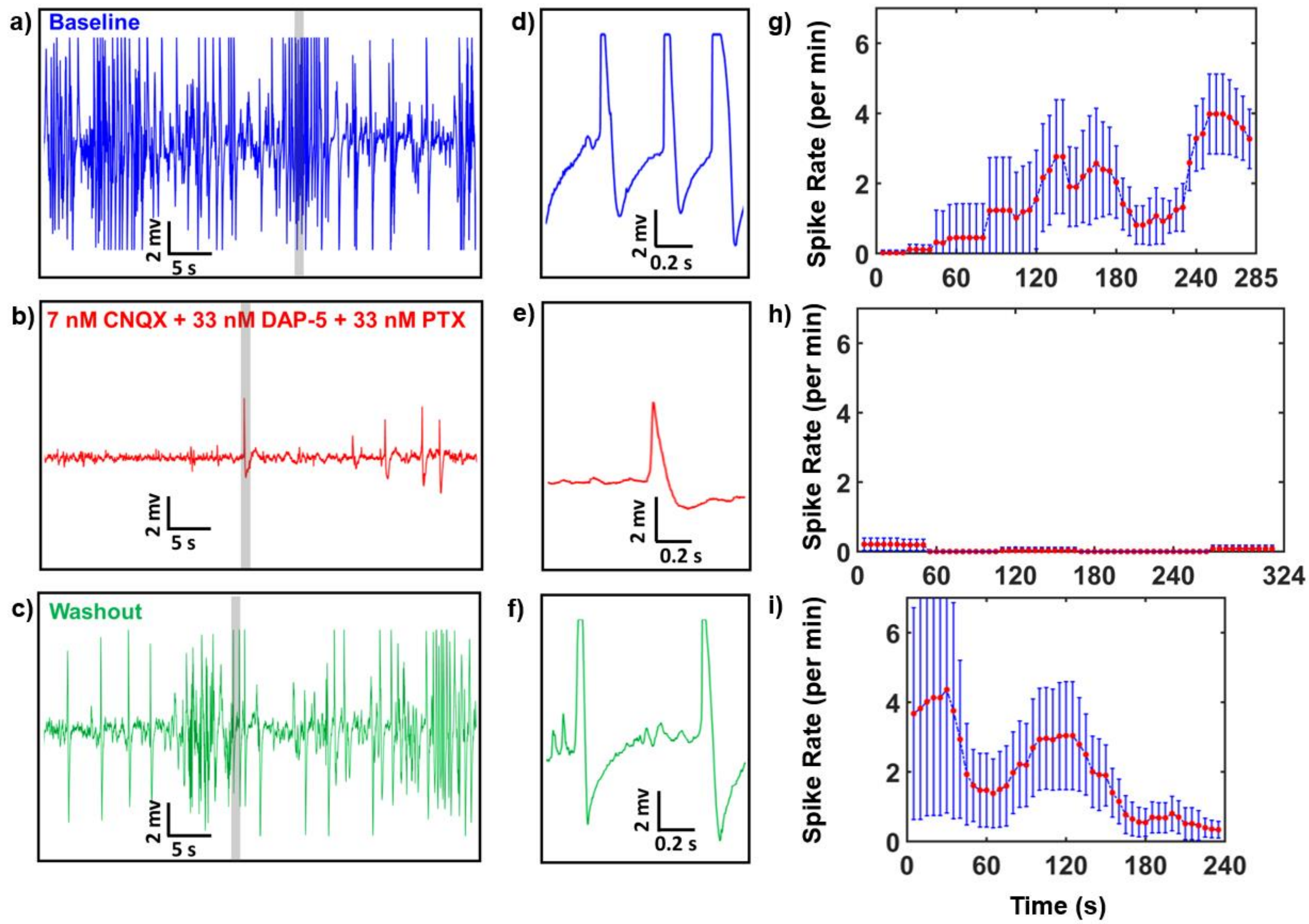

**Figure S21.** Example pharmacological interrogation at 7 DIV involving CNQX, DAP-5, and PTX to validate the pre-potentials. Expanded views the recording traces at the gray-highlighted regions are shown. a) 50 s baseline recording trace with action potential spikes. b) 50 s recording trace after addition of 7 nM CNQX, 33 nM DAP-5, and 33 nM PTX. c) 50 s recording trace after sequential washout. d) Zoomed-in 1 s segment of baseline recording. e) Zoomed-in 1 s recording segment after addition of 7 nM CNQX, 33 nM DAP-5, and 33 nM PTX. f) Zoomed-in 1 s recording segment after washout. g) Spike rate per min of baseline recording (285 s total) after high pass filtering h) Spike rate per min of recording after addition of 7 nM CNQX, 33 nM DAP-5, and 33 nM PTX (324 s total) after high pass filtering i) Spike rate per min of recording after washout (240 s total) after high pass filtering.

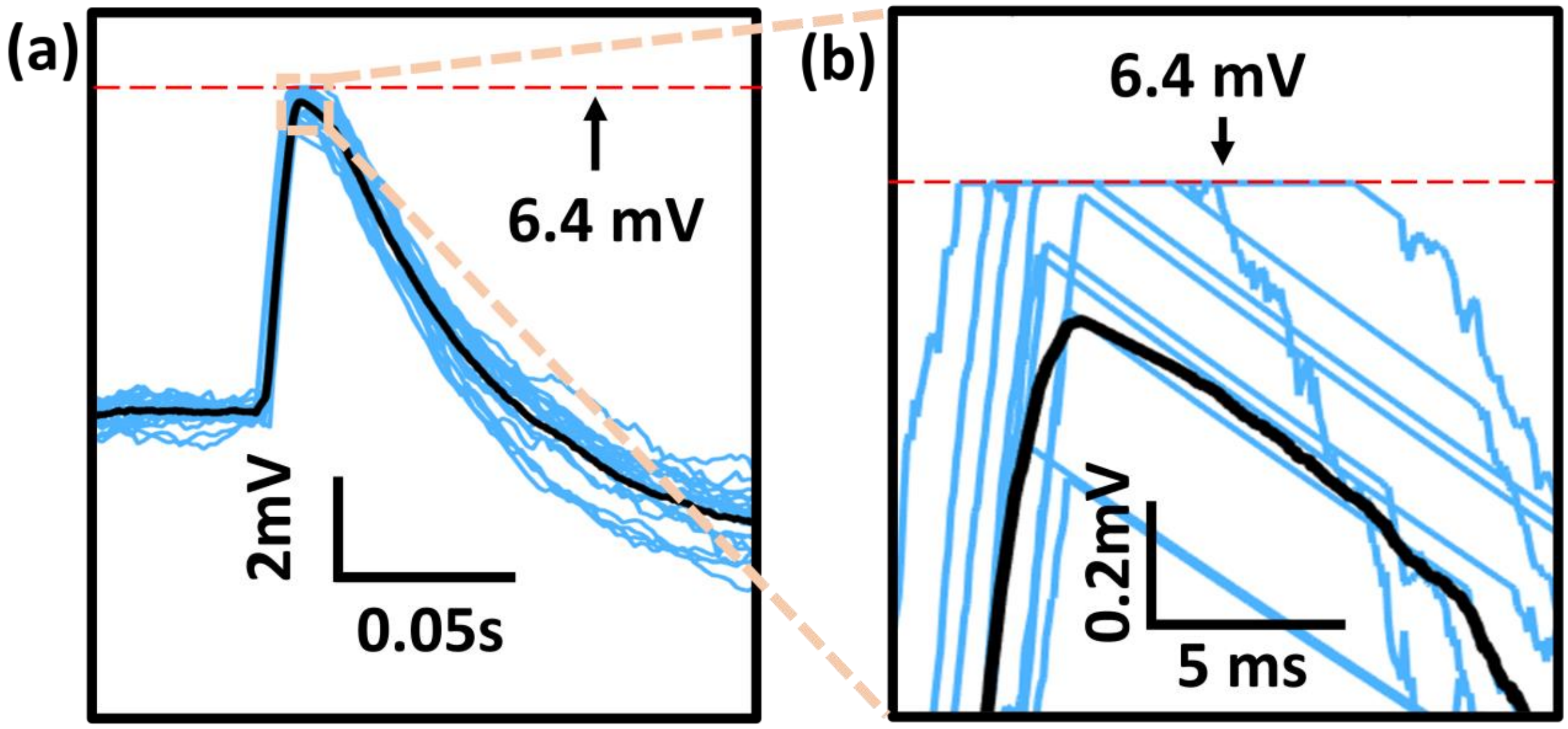

**Figure S22.** a-b) Selected time segment of the spike-sorting result shows clipping of recorded action potentials of *in vitro* iPSC-CVPCs at the Intan amplifier analog-to-digital converter (ADC) limit of +/- 6.4 mV.

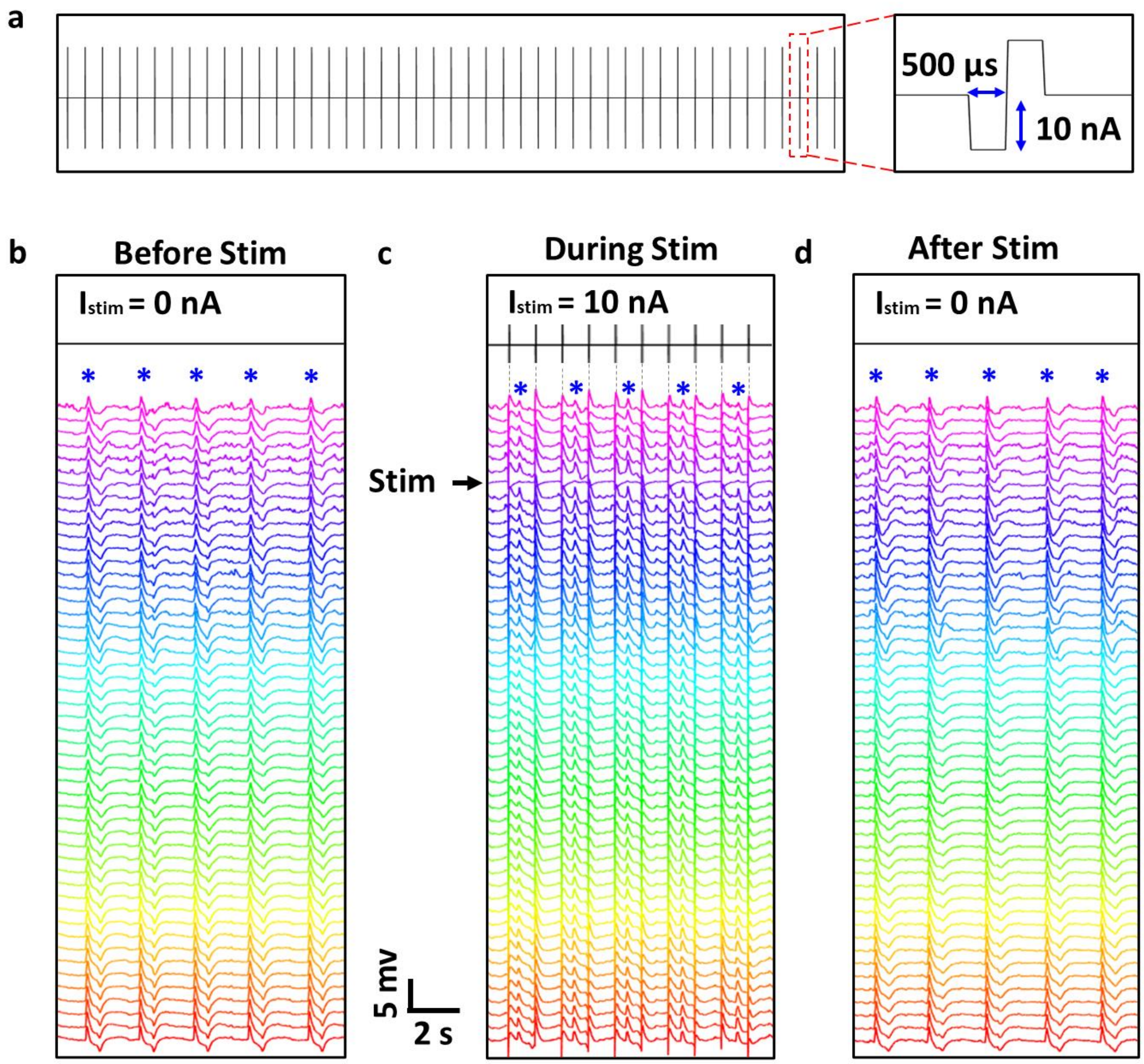

**Figure S23.** a) The biphasic-pulse stimulation traces with stimulation peak width, amplitude and frequency of 0.5 ms, 10 nA and 1 Hz, respectively. b) 52-Channel voltage traces of 2 arrays recorded from the iPSC-CVPCs at day 34 of differentiation before stimulation. c) 52-channel voltage traces of 2 arrays recorded from the iPSC-CVPCs at day 34 of differentiation during stimulation with the stimulation channel labeled in c). Gray dashed lines align the stimulation peaks with "artifact" spikes recorded from USNW electrodes due to capacitive coupling. d) 52-Channel voltage traces of 2 arrays recorded from the iPSC-CVPCs at day 34 of differentiation after stimulation. Action potential peaks are labeled as blue stars in b-d).

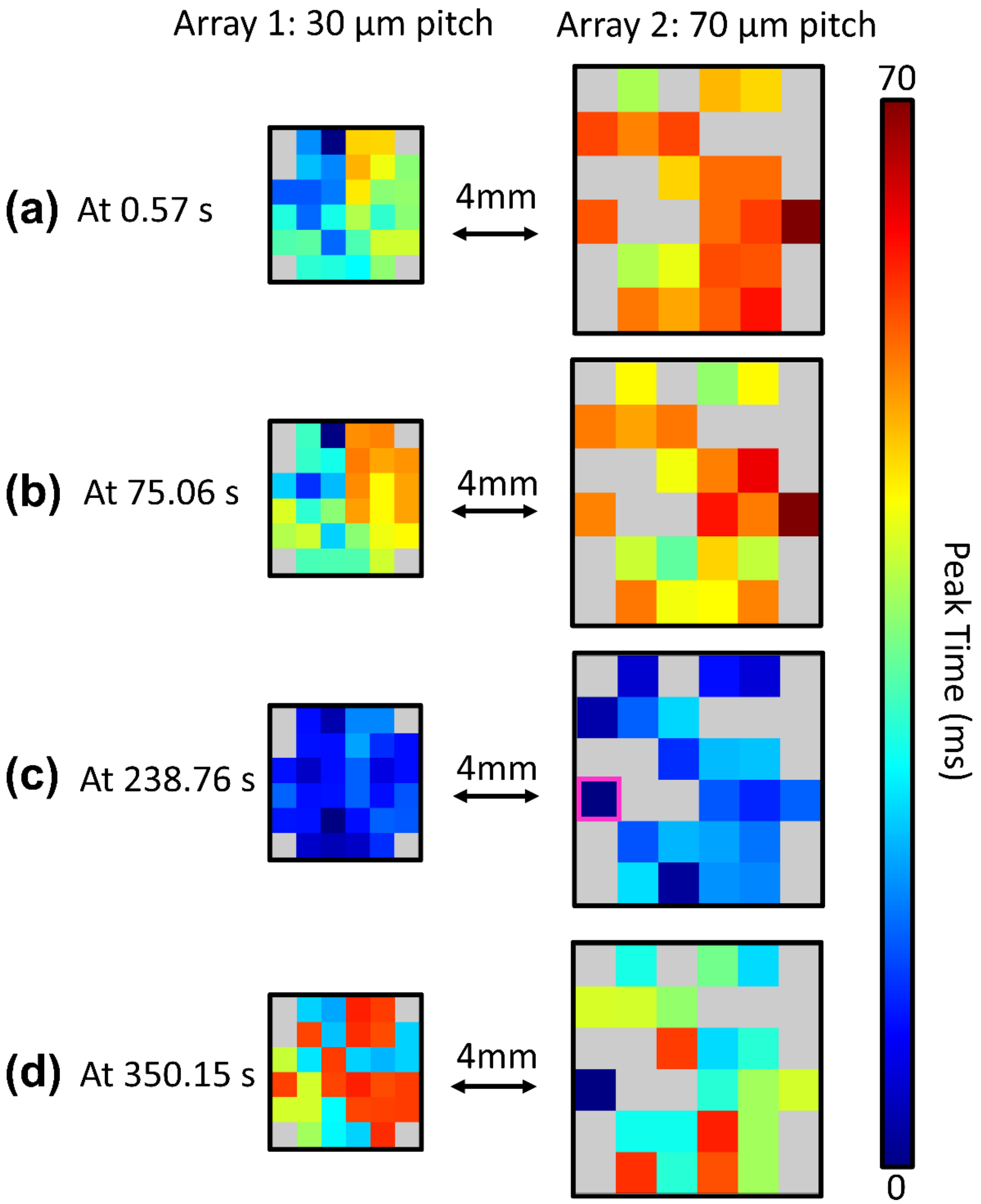

**Figure S24.** Mapping of action potential propagation patterns across the 2 arrays according to the array size scale at different time points before (a) at 0.57 s and b) at 75.06 s) and after (c) at 238.76 s and d) at 350.15 s) electrical stimulation. The distance between array 1 with 30 µm pitch and array 2 with 70 µm pitch was 4 mm. Two intracellular recordings before electrical stimulation (at 0.57 s and at 75.06 s) show action potential propagation from left to right, whereas intracellular recordings after electrical stimulation show an evolution from homogeneous firing at 238.76 s to a reserved propagation direction from right to left, where the action potential propagation direction starts from the simulating electrode. The biphasic-pulse stimulation peak width, amplitude and frequency were 0.5 ms, 10 nA and 1 Hz, respectively. The stimulation channel location is shown as the pink box in c).

**6. Channel selection, auto-thresholding spike sorting, and detected spike data analysis**

6.1 Channel selection via cross-correlation matrix

For signal amplitude and spike sorting data analysis, observed channels and chosen USNW bank arrays were selected based on the USNW group and channels exhibiting largest amplitudes and displayed intracellular action potential waveforms upon visual inspection. For recording datasets with relatively low intracellular activities 11 DIV for single USNW, 11 to 19 DIV for 16 USNW and 625 USNW, channels and groups exhibiting low signal cross-correlation coefficient between one another (Figure S25i-k) to isolate channels with electrophysiological signals. The cross-correlation coefficient plot (Figure S25e-k) between each channel were determined based on sample 10 s data segments, with darker colors representing lower correlation. Figure S25e,f show cross-correlation plot of group B channels (from 11 to 19 DIV) and show evolution of neural networks over the culture time. Figure S25a-d show the corresponding longest segment from group B array for 11 to 19 DIV dataset. Primarily, group B array and few channels from group C array for 625 USNWs per channel sample that were incorporated into the signal amplitude analysis. For pharmacological analysis (at 7 DIV), single channel of USNWs were chosen and corresponding spike rates were calculated and observed.

Starting from 13 DIV, we observed strongly correlated activity across all channels within an individual array. The measured signals were essentially identical across majority of the channels (extremely similar spike counts as shown in Figure 2c,e) and showed strong inter-channel cross-correlation (Figure S25f,g). These most likely resulted since the time delay is too short to be detected because of the high density/tight pitch of our USNW electrodes for recording the signal propagation of the neurons in the network. According to the published result that the propagation speed for *in vitro* neuronal network is 110 mm $s^{-1}$,[8] the time delay between two electrodes with 10 μm spacing is less than 0.1 ms, which is nearly the limit for our recording temporal resolution.

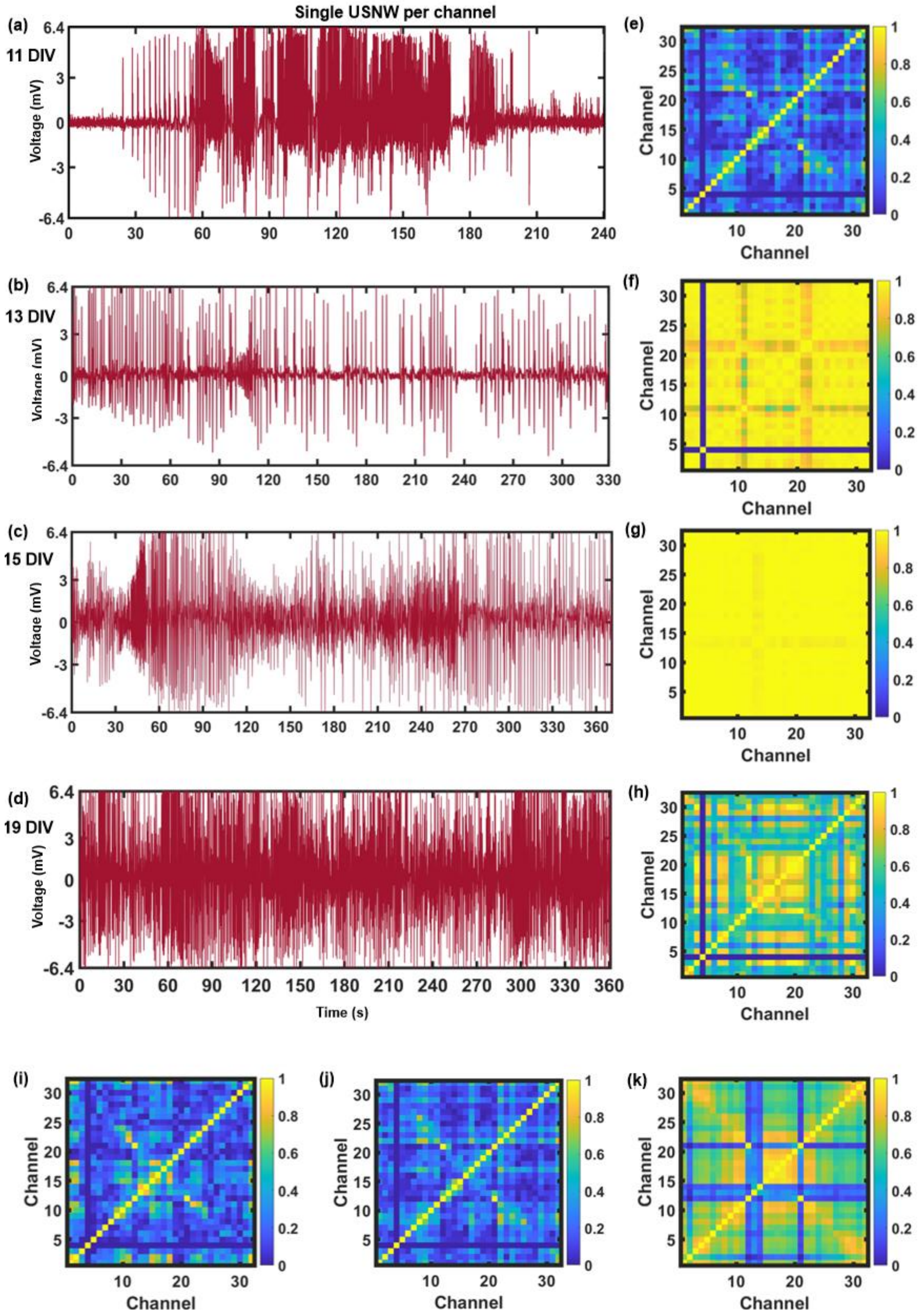

**Figure S25.** Longest single USNW recording segments 11-19 DIV and corresponding spike counts and spike rates a) Continuous, 240 s recording segment from channel 31 USNW from group B (10 µm electrode spacing) at 11 DIV (spike rate of 1.89 spikes $s^{-1}$). b) Continuous, 330 s recording segment from channel 31 USNW from group B at 13 DIV (mean spike rate of 0.45 spikes $s^{-1}$ across 31 channels). c) Continuous, 370 s recording segment from channel 31 USNW from group B at 15 DIV (mean spike rate of 1.13 spikes $s^{-1}$ across 32 channels). d**)** Continuous, 360 s recording segment from channel 31 USNW from group B at 19 DIV (mean spike rate of 2.52 spikes $s^{-1}$ across 31 channels). e-h) Cross-correlation matrix for selected USNW groups for analysis from 11, 13, 15, and 19 DIV. e) Single USNW per channel, group B at 11 DIV. f) Single USNW per channel, group B at 13 DIV. g) Single USNW per channel, group B at 15 DIV. h) Single USNW per channel, group B at 19 DIV. The low cross-correlation correspond to large variance in spike counts observed for Figure 2g. i-k) Example cross-correlation matrix for selected USNW groups for analysis from batches at 11 DIV, i) 16 USNWs per channel, group B at 11 DIV, j) 625 USNWs per channel, group B at 11 DIV, k) 625 USNWs per channel, group C at 11 DIV.

6.2 Auto-threshold spike sorting

For spike sorting, selected channels for data analysis were high-pass filtered at 270 Hz for pharmacological experiments and unfiltered for signal amplitude comparison studies to retain the recorded peak-to-peak amplitude (to maximize the amplitude fidelity and perform more impartial comparison between pads driving different USNW counts). The positive and negative threshold values were determined and set for every selected channel with multiple of the normalized standard deviation, where our estimated thresholds were set to range from 4 – 6.5 $\sigma_n$ [9] for high-pass filtered data and thresholds were set and ranged from 1 to 4 $\sigma_n$ for unfiltered, raw data. After determining the positive and negative threshold values from the autothresholding algorithm for every selected channel, spike sorting was carried out for all channels with the respective threshold values against data segments recorded after biphasic current injection (data segment length of 180 s for pharmacology experiments and approximately 240 to 370 s for signal amplitude studies from recording segments at 11 to 19 DIV).

Spikes of both polarities, exceeding both positive and negative thresholds with temporal widths between 5 ms and 100 ms of each other (respectively considering the short spikelets and temporal broadening of potential spikes from the parasitic capacitance of the measurement system), were coupled as one spiking event to ensure intracellular action potentials were detected. Our spike detection begins when the signal initially exceeds the positive threshold. The local maximum is then determined between the time segment when the signal first exceeds and decreases back down to the set positive threshold again. From the local, positive maximum, if the spikes reached negative threshold within reasonable timeframe (within 100 ms) without the signal exceeding positive threshold for longer than 5 ms, then the local minimum is then determined between the time segment when the signal exceeds and reaches the set negative threshold again. Peak-to-peak amplitude is the magnitude difference between the local maximum and local minimum. Figure S26 shows an example segment of our spike detection method for unfiltered signal with the thresholds determined from the auto-threshold algorithm. By defining spikes to always contain positive and negative spike components alongside reasonable time duration between such two spikes, short, low-amplitude, and unipolar spikelets exceeding their corresponding positive or negative thresholds were automatically excluded out during the spike sorting.

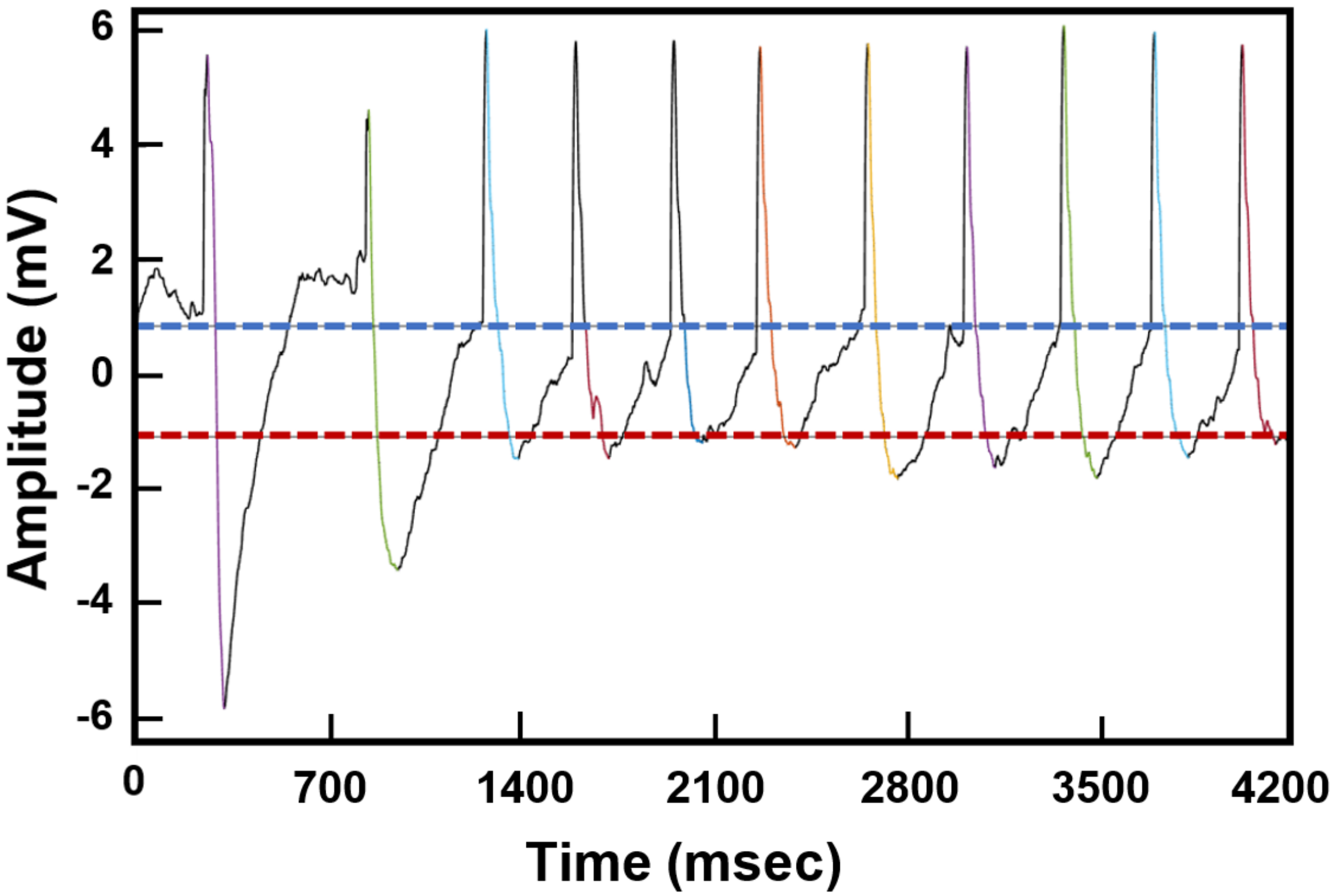

**Figure S26.** Auto-thresholding spike sorting example plot with positive and negative thresholds a) Example of unfiltered signal and detected spikes with positive and negative thresholds (dashed blue and red lines respectively, thresholds are set at 1.6 and -1.6 σ). Peak-to-peak amplitude is determined from local maxima and minima respectively above and below the thresholds, shown with varied colored lines.

6.3 Misc. data analysis from detected spikes

For spike rate analysis from 180 s time segment (Figure 2p-r) for baseline signal, signal after PTX application, and signal after TTX application, spike rates were calculated by determining the number of detected spikes every 5 s. With spike rate determined for every 5 s, 35 spike rate data points were plotted for selected channels over 180 s of recording. Channels exhibiting large, more than tenfold, deviation from the mean spike rate were excluded (ultimately, 27 channels were used to plot the spike rates for baseline signals, signals with PTX, and signals with TTX).

Interspike interval is determined by measuring the time between each positive, maximum peaks of the detected spikes, and the obtained interspike intervals are centered around 500 to 700 ms range (Figure 3g). The spike widths range near approximate mid-point of the recorded intracellular potentials range up to approximately 150 ms and were generally around 50 to 75 ms (Figure 2l), so the calculation of interspike interval based on the spikes' maximum peak point is a fair approximation.

**7. Small Signal Circuit Modeling of Individually Addressable USNW versus Multi-USNWs per Channel**

Building off of the circuit modeling of the electro-neural interface we performed previously, [10] we simulated the effects of adding multiple NWs on a single recording channel. We started by adopting a typical patch-clamp measured sodium and potassium transmembrane current as a signal source, and building an appropriate RC network around this source. Figure 4h showed this small signal model for a single intracellular USNW penetrating a neural cell body. We modeled the excitable cell as having two lumped membrane impedances, one of which contains active ion channels that contribute to the generation of an action potential, denoted as a junctional membrane, and the other which is assumed to be a passive impedance modeling leakage from the inside of the cell to the outside of the cell, denoted as a non-junctional membrane. Simulations were carried out using CADENCE Spectre circuit simulator software, and values for membrane capacitance, ion channel conductance, and spreading resistance were all matched to achieve an intracellular potential waveform shown in Figure S27 below. The ground of this circuit model exists some distance away from the cell; thus there exists some non-zero electrical impedance between the outside of the cell and ground. We modeled this impedance as a spreading resistance ($R_{spread}$) of electrolytic solution, and thus, a non-zero potential can exist just outside of the cell during an action potential. Note that this is important for simulations of extracellular or juxtacellular configurations where the NWs do not fully penetrate the cell membrane. Further, the electrochemical interface between the intracellular fluid and the intracellular USNW is modeled as a simple parallel RC circuit ($R_{EC}$, $C_{EC}$) which aims to capture both faradaic and capacitive electrochemical transduction processes, and the quality of the membrane seal around this penetrating electrode is modeled with a sealing resistance ($R_{seal}$). Finally, depending on the cell geometry, extracellular fluid conductance, and other factors, there may exist a signal pathway between the non-junctional extracellular region and the junctional extracellular region; we introduced here an isolation resistance ($R_{iso}$) which attempts to capture this effect.

**(a)**

| Parameter | Value |
|---|---|
| Camp | 5.5pF |
| Cec | 1pF |
| Cmj | 10fF |
| Cmnj | 1fF |
| Cp | 1pF |
| Rec | 65Mohm |
| Riso | 8kohm |
| Rmj | 200Mohm |
| Rmnj | 10Gohm |
| Rseal_extra | 1kohm |
| Rseal_intra | 10Gohm |
| Rspread | 10kohm |

**(b)**

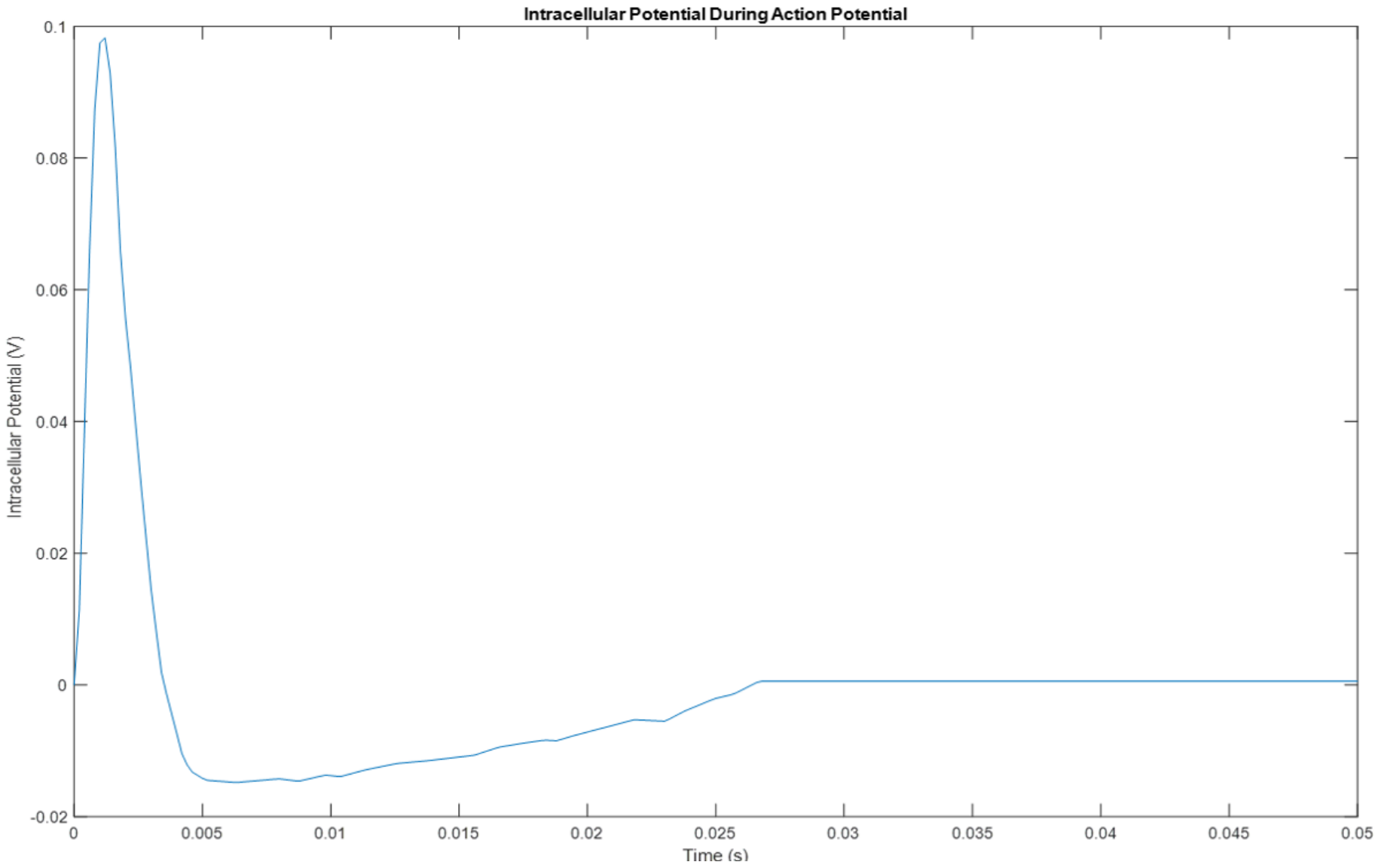

**Figure S27.** Default circuit parameters and corresponding simulated intracellular potential without any NW connected. The intracellular potential resulted from transmembrane currents taken from [10]. Cell circuit parameter values were tuned to achieve an intracellular to known amplitudes and spike durations of ~100mV peak with ~4 to 5 ms positive phase duration.

**Uncategorized References**